\documentclass[11pt]{article}
\usepackage[dvips]{graphicx}
\usepackage{bm}
\usepackage{amsmath}
\usepackage{amssymb}
\usepackage{float}
\usepackage{url}
\renewcommand{\baselinestretch}{1.1}
\newcommand{\nn}{\nonumber\\}

\newcommand{\ket}[1]{\left| #1 \right>}
\renewcommand{\thepage}{}
\makeatletter
\@addtoreset{equation}{section}
\renewcommand{\theequation}{\thesection.\@arabic\c@equation}
\makeatother
\renewcommand{\thefootnote}{\fnsymbol{footnote}}
\begin{document}
\begin{titlepage}
\title{
\vspace*{-4ex}
\hfill{}\\
\hfill
\begin{minipage}{3.5cm}
\end{minipage}\\
\bf Closed string symmetries
 in open string field theory
\vspace{1ex}\\
{\Large
\it -- tachyon vacuum as sine-square deformation --}
\vspace{3ex}
}

\author{
\ \ Isao {\sc Kishimoto}$^{1}$\footnote{ikishimo@ed.niigata-u.ac.jp},
\ \ Tomomi {\sc Kitade}$^{2}$\footnote{kitade@asuka.phys.nara-wu.ac.jp}
\ \ and\ \  
Tomohiko {\sc Takahashi}$^{2}$\footnote{tomo@asuka.phys.nara-wu.ac.jp}
\\
\vspace{0.5ex}\\
\\
$^{1}${\it Faculty of Education, Niigata University,}\\
{\it Niigata 950-2181, Japan}
\vspace{1ex}
\\
$^{2}${\it Department of Physics, Nara Women's University,}\\
{\it Nara 630-8506, Japan}
\vspace{2ex}
}

\date{}
\maketitle

%
\vspace{7ex}

\begin{abstract}
\normalsize
We revisit the identity-based solutions for tachyon condensation in open
bosonic string field theory (SFT) from the viewpoint of the sine-square
 deformation (SSD).
The string Hamiltonian derived from the simplest solution
includes the sine-square factor, which is the same as that of an open system 
with SSD in the context of condensed matter physics.
We show that the open string system with SSD or its generalization exhibits
decoupling of the left and right moving modes and so
it behaves like a system with a periodic boundary condition.
With a method developed by Ishibashi and Tada,
we construct pairs of Virasoro generators in this system,
which represent symmetries for a closed string system.
Moreover, we find that 
the modified BRST operator in the open SFT at the
identity-based tachyon vacuum decomposes to holomorphic and
antiholomorphic parts, and these
reflect closed string symmetries in the open SFT.
On the
basis of SSD and these decomposed operators, we construct
holomorphic and antiholomorphic continuous Virasoro algebras at the
 tachyon vacuum.
These results imply
that it is possible to formulate a pure
closed string theory in terms of the open SFT at the identity-based
tachyon vacuum.
\end{abstract}

\vspace{5ex}

\end{titlepage}

\renewcommand{\thepage}{\arabic{page}}
\renewcommand{\thefootnote}{\arabic{footnote}}
\setcounter{page}{1}
\setcounter{footnote}{0}
%
\renewcommand{\baselinestretch}{1}
\tableofcontents
\renewcommand{\baselinestretch}{1.1}

\newpage
\section{Introduction}

Open bosonic string field theories on a D-brane have a stable vacuum
where a tachyon field is condensed and the D-brane disappears.
At the tachyon vacuum, it is proved that we
have no open strings as physical degrees of freedom and
there is a belief that closed strings living in a bulk space-time
are found.
Since the on-shell closed string spectrum is obtained in a scattering
amplitude in the perturbative vacuum of open string field theories, it
is reasonable that there are some observables made by open
string fields to clarify the existence of closed strings even on the
tachyon vacuum. If it was realized, a pure closed string theory would be
formulated in terms of open string fields.

In string theories, it has long been considered that closed string
dynamics is described by open string degrees of freedom.
It is naively intuitive that a closed string is given by a bound state of
an open string and so an open string is regarded as a fundamental
degree of freedom. This idea is realized in matrix theories,
which are comprehensive or constructive theories including closed
strings and D-branes, where matrices are
 considered as collections of
zero modes of open strings.
The AdS/CFT correspondence also provides a sophisticated version of
this thought.

About a decade ago, Gendiar, Krcmar and Nishino made an important
discovery concerning discretized open systems in condensed matter
physics \cite{GKN}. 
As an example,
they considered an $N$-site system with open boundary condition described 
by the Hamiltonian
\begin{eqnarray}
 \hat{H}=-t\sum_{l=1}^{N-1}\sin^2\frac{l\pi}{N}
\left(\hat{c}_l^\dagger \hat{c}_{l+1}+\hat{c}_{l+1}^\dagger
\hat{c}_l\right),
\label{eq:ssdH}
\end{eqnarray}
where $\hat{c}_l$ denotes a spinless free fermion on a 1D
lattice, which satisfies $\{\hat{c}_l,\,\hat{c}_{l'}^{\dagger}\}=\delta_{l,l'}$.
This is a deformed Hamiltonian of a uniform open system by the
sine square factor.
They found that this system is equivalent to a system with a periodic
boundary condition by calculating the ground state energy and correlation
functions numerically. 
This result suggests that a discretized closed system can be described
by degrees of freedom of a discretized open system, and, considering the
continuum limit, the above idea is realized in a 2D field theory.

After their seminal papers \cite{GKN}, the deformation such as
(\ref{eq:ssdH}) is called the sine-square deformation (SSD) and it has
been actively studied. The above correspondence between the open and
closed systems is understood as a
result of the equivalence of the ground states of the SSD and uniform
periodic systems \cite{HN}, and after that it has been proved
exactly \cite{KatsuraEGS}. The SSD has been extended to
deformation with other suppression of boundary effects and the generalized
SSD has been applied not only to the free fermion system but also to
various models \cite{GKN,HN,KatsuraEGS,SH,MKH}. 
Moreover, SSD has also been studied in the context of a field theoretical
description \cite{Katsura:2011ss, Tada:2014wta, Tada:2014kza,
Ishibashi:2015jba, Ishibashi:2016bey,
Okunishi:2016zat, Wen:2016inm}.
These studies strongly suggest the equivalence of an SSD system and a
periodic system, and, from the string theoretical viewpoint,
the possibility of closed string theories in terms of open strings.

Interestingly, it has been pointed out in \cite{Ishibashi:2016bey} that
there is an intimate
relation between the SSD system and open string field theories.
On the string field theoretical side, it is known that one of the
identity-based tachyon vacuum solutions \cite{Takahashi:2002ez,Kishimoto:2002xi} 
provides a kinetic operator corresponding to a string Hamiltonian at the tachyon 
vacuum \cite{Takahashi:2003xe}:
\begin{eqnarray}
 L'=\frac{1}{2}L_0'-\frac{1}{4}(L_2'+L_{-2}')+\frac{3}{2},
\end{eqnarray}
where $L_n'$ is the twisted ghost Virasoro operator with the central
charge 24. 
This operator
can be rewritten by using the twisted
ghost energy-momentum tensor $T'(z)$:
\begin{eqnarray}
 L'=\oint \frac{dz}{2\pi i}\frac{-1}{4z}\left(z^2-1\right)^2 \left(
T'(z)-6\right).
\end{eqnarray}
If we change the variable to $z=\exp(i\theta)$, the weighting
function in $L'$ becomes $\sin^2\theta$
due to a Jacobian and a conformal factor
from $T'(z)$. Hence, the kinetic operator turns out to be related
to an SSD Hamiltonian in conformal field theories
\cite{Ishibashi:2016bey}.

The identity-based solution is constructed by an integration of some
local operators along a half string and the identity string field.
As a result, the theory expanded around the solution includes the
modified BRST operator expressed as the integration of local operators
on a worldsheet, which is a useful tool
for investigating the corresponding vacuum.
Actually,
we can find the perturbative
vacuum in the theory expanded around the
identity-based tachyon vacuum solution with high precision by a level truncation
scheme \cite{Takahashi:2003ppa,Kishimoto:2009nd,Kishimoto:2011zza},
although the vacuum energy is analytically calculated
later \cite{Kishimoto:2014lua, Ishibashi:2014mua, Zeze:2014qha}.
This is a remarkable feature that has never been seen in wedge-based
solutions \cite{Schnabl:2005gv, Erler:2009uj}.
Furthermore, 
closed string amplitudes generated by
a propagator on the identity-based tachyon vacuum 
have been formally discussed \cite{Drukker:2003hh, Drukker:2002ct}
and then a vacuum loop amplitude generated by the propagator was
numerically calculated \cite{Takahashi:2011zzb}.

These facts about SSD and string field theories (SFT) indicate that
we are able to formulate a pure closed string theory 
in terms of open string fields and the identity-based tachyon vacuum solution.
In this paper, motivated by this possibility, 
we will try to find closed string symmetries in the open
string field theory 
at the identity-based tachyon vacuum.
To do so, we will mainly use
the technique developed in dipolar quantization of 2D 
conformal field theories \cite{Tada:2014wta, Tada:2014kza, Ishibashi:2015jba, Ishibashi:2016bey}.
Consequently, we will obtain holomorphic and antiholomorphic continuous
Virasoro operators that commute with the modified
BRST operator on the identity-based tachyon vacuum.
The existence of such Virasoro operators implies that open SFT on the
identity-based tachyon vacuum provides some observables given by the
worldsheet without boundaries, i.e., observables of closed string
theories.

This paper is organized as follows. In section \ref{sec:SSLD}, 
we will discuss sine-square-like deformation (SSLD) of open string systems. 
With the help of the
dipolar quantization technique, we will explain how left and right
moving modes, which correlate in an open boundary system, become
decoupled in the system with SSLD.
Due to this decoupling, we can find left and right continuous Virasoro
algebras, namely holomorphic and antiholomorphic parts of the Virasoro
algebra.
Here, we emphasize that this left-right decoupling is the most important
ingredient for realizing the equivalence between
the system with SSLD and that with a periodic boundary condition. 
In section \ref{sec:CSS_TV}, we will discuss closed string symmetries in open string
field theories at the identity-based tachyon vacuum. After a brief review
of the identity-based tachyon vacuum solution, we will show that the
modified BRST operator decomposes to the holomorphic and antiholomorphic
parts and that they are nilpotent and anticommute with each other.
It will be shown that these correspond to the holomorphic and
antiholomorphic parts of the BRST operator of a closed string.
Next, we will construct the energy-momentum tensor and Virasoro operators at
the tachyon vacuum, which are related to closed string symmetries in the
open string field theory. It is noted that the energy-momentum tensor
given in this section includes a new notion of frame-dependent
operators, and it is not the twisted ghost operators that appeared
previously. Then, we will discuss ghost operators and ghost number at
the tachyon vacuum. In addition, we will comment on other types of 
identity-based tachyon vacuum solutions. In section \ref{sec:Conclusion}, we will give
concluding remarks. Details of calculations are provided in the appendices.

\section{Sine-square-like deformations of an open string system
\label{sec:SSLD}
}
\subsection{Decoupling of left and right moving modes
\label{sec:decoupling}}

The Hamiltonian of an open string is given by
\begin{eqnarray}
 H_{\rm O}=\int_{C_+}\frac{dz}{2\pi i}\,z T(z)
+\int_{C_-}\frac{dz}{2\pi i}\,z T(z),
\label{eq:Ho}
\end{eqnarray}
where $T(z)$ is the energy-momentum tensor, $C_+$ ($C_-$) denotes a
counterclockwise contour along a unit semicircle in a upper (lower) half plane.
The end points of $C_\pm$, $z=\pm 1$, correspond to open string
boundaries.
The first and second terms correspond to Hamiltonians of the left and
right moving modes respectively, but
they do not commute with each other due to open boundary conditions on
$T(z)$.
Since the two contours $C_+$ and $C_-$ form a circle, the
Hamiltonian is given by the zeroth component of the Virasoro operators
$L_n$.
Thus, we do not encounter antiholomorphic Virasoro operators in the
open string system.

Here, we consider the deformation of (\ref{eq:Ho}) by introducing
a weighting function $g(z)$:
\begin{eqnarray}
 H_g=H_g^+ + H_g^-,
~~~~~~
H_g^\pm=\int_{C_\pm}\frac{dz}{2\pi i}\,g(z) T(z),
\label{eq:Hg}
\end{eqnarray}
where $g(z)$ is a holomorphic function satisfying $g(\pm 1)=\partial
g(\pm 1)=0$ and 
$\left(g(1/z^*)\right)^*=g(z)/z^2$
is imposed for the hermiticity
of $H_g$. 
$H_g^+$ and $H_g^-$ are left and right moving modes of $H_g$.
The simplest example of $g(z)$ is given by
\begin{eqnarray}
 g(z)=-\frac{1}{4z}(z^2-1)^2.
\label{eq:gsin2}
\end{eqnarray}
As mentioned in the introduction, if we change the variable to
$z=\exp(i\theta)$, the weighting
function in $H_g$ is changed to $z^{-1} g(z)=\sin^2\theta$.
Hence, the deformed Hamiltonian (\ref{eq:Hg})
provides a sort of generalization of the SSD Hamiltonian.
In this sense, we call it the sine-square-like deformation,
or SSLD for short.

Since $T(z)$ in (\ref{eq:Hg}) is the same that as in (\ref{eq:Ho}), 
it is expanded by holomorphic Virasoro operators only:
\begin{eqnarray}
 T(z)=\sum_{n=-\infty}^\infty L_n z^{-n-2}.
\end{eqnarray}
By using this expansion form and the Virasoro algebra, we can obtain a
commutation relation for $T(z)$ (appendix \ref{sec:cr_SSLD}):
\begin{eqnarray}
 [T(z),\,T(z')]=-(\,T(z)+T(z')\,)\,\partial
  \delta(z,z')-\frac{c}{12}\partial^3 \delta(z,z'),
\label{eq:TT}
\end{eqnarray}
where $c$ is the central charge of $T(z)$ and the delta function
$\delta(z,z^{\prime})=\sum_{n=-\infty}^{\infty}z^{-n}z^{\prime n-1}$ is normalized 
as \cite{Takahashi:2002ez}
\begin{eqnarray}
 \int_{C_++C_-}\frac{dz}{2\pi i}\,\delta(z,z')=1.
 \label{eq:delta_normalization}
\end{eqnarray}

By using (\ref{eq:TT}), we can calculate the
commutation relation between $H_g^+$ and $H_g^-$.
The important point is that surface terms appear in the calculation as
a result of derivatives of the delta function
and these terms include a singular factor $\delta(\pm 1,\pm 1)$.
However, the singular surface terms turn out to vanish due to the
factors $g(\pm 1)$ and $\partial g(\pm 1)$, which are set to zero in the
definition of $H_g$.
As a result, we find that $H_g^{+}$ commutes with $H_g^{-}$  (\ref{eq:app_Hg+Hg-=0})
and then the deformed system is decomposed
into left and right moving parts as in periodic systems. 
Accordingly, it is concluded that the deformed system described by $H_g$
corresponds not to an open string system, but to a closed string system,
although the Hamiltonian is constructed by a single holomorphic
energy-momentum tensor.

It should be noted that the zeros of $g(z)$ and $\partial g(z)$ at open
string boundaries cause decoupling of the left and right moving sectors.
This is the mechanism by which periodic properties are achieved in the
SSLD system.

\subsection{Example of string propagations
\label{sec:example}}

Equal-time contours generated by $H_g$ have been discussed by
Ishibashi and Tada in \cite{Ishibashi:2016bey}. However, they did not
explicitly consider left and right moving modes.
In this subsection, we will illustrate equal-time contours generated by the
Hamiltonian for the simplest function (\ref{eq:gsin2})
with a focus on the emergence of left and right moving sectors, 
although it corresponds to 
the $Z_2$-symmetric case in \cite{Ishibashi:2016bey}.

According to \cite{Ishibashi:2016bey}, we introduce the parameters, $t$
and $s$, into the worldsheet generated by $H_g$ for the function
(\ref{eq:gsin2}):
\begin{eqnarray}
 t+i s=\int^z \frac{dz}{g(z)}=\frac{2}{z^2-1},
\label{eq:ts}
\end{eqnarray}
where $t$ denotes time and $s$ parametrizes a string at a certain time.
If we introduce polar coordinates as $z=r e^{i\theta}$, they can be
expressed by the parameters $t$ and $s$ on the upper half plane (${\rm
Im}\, z\geq 0$),
\begin{eqnarray}
 \theta&=&\left\{
\begin{array}{ll}
\displaystyle
 \frac{\pi}{4}+\frac{1}{2}\arctan\frac{s^2+2t
  +t^2}{2s} & (s< 0)\\
&\\
\displaystyle
 \frac{3\pi}{4}+\frac{1}{2}\arctan\frac{s^2+2t
  +t^2}{2s} & (s>0),
\end{array}\right.
\label{eq:thetaplus}
\\
r&=& \left(\frac{s^2+(t+2)^2}{s^2+t^2}\right)^{\frac{1}{4}},
\label{eq:radius1}
\end{eqnarray}
and
on the lower half plane (${\rm
Im}\, z\leq 0$),
\begin{eqnarray}
 \theta&=&\left\{
\begin{array}{ll}
\displaystyle
 -\frac{\pi}{4}+\frac{1}{2}\arctan\frac{s^2+2t
  +t^2}{2s} & (s>0)\\
&\\
\displaystyle
 -\frac{3\pi}{4}+\frac{1}{2}\arctan\frac{s^2+2t
  +t^2}{2s} & (s<0),
\end{array}\right.
\label{eq:thetaminus}
\\
r&=& \left(\frac{s^2+(t+2)^2}{s^2+t^2}\right)^{\frac{1}{4}}.
\label{eq:radius2}
\end{eqnarray}
The contour at the time $t=-1$ is given by the unit circle $r=1$, in
which $s$ is related to $\theta$ as $s=-\cot \theta$. As $\theta$
increases from $0\ (z=1)$ to $\pi\ (z=-1)$ on the upper half
plane, $s$ increases from $-\infty$ to $\infty$. Similarly, on the lower
half plane, $s$ increases from $-\infty$ to $\infty$ 
with increasing $\theta$ from $-\pi$ to $0$. 

\begin{figure}
\begin{center}
\includegraphics[width=16cm]{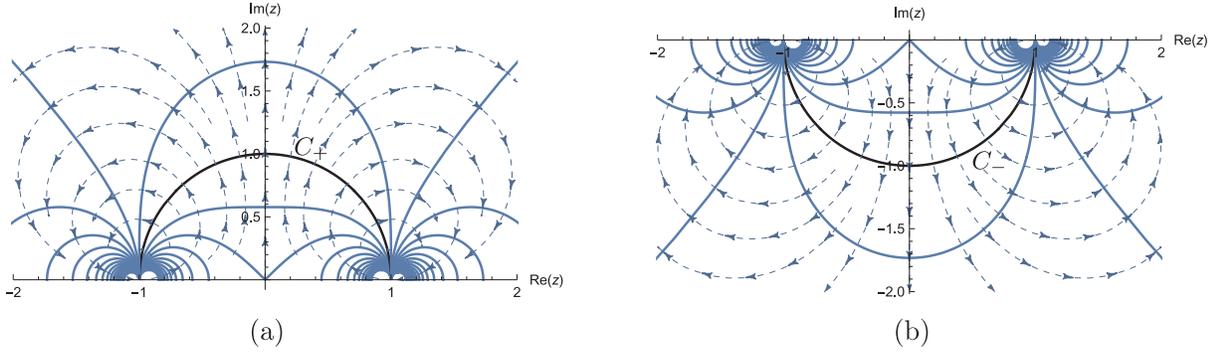}
\end{center}
\caption{Equal-time contours on the $z$ plane (solid lines). Dashed lines with arrows denote evolution of time $t$.}
\label{fig:contour} 
\end{figure}

The upper half plane corresponds to the worldsheet swept by the left moving
 sector of a string. 
Equal-time contours can be depicted by using (\ref{eq:thetaplus}) and
(\ref{eq:radius1}), in which $s$ runs from $-\infty$ to $\infty$ as in the
case of $t=-1$.
As the equal-time contour in
Fig.~\ref{fig:contour} (a), the unit semicircle $C_+$ at $t=-1$ becomes
larger as 
time increases, while the boundaries of a string do not propagate due
to the zeros of $g(z)$. At $t=0$, the midpoint of the string goes
to infinity. After that the midpoint splits into two points on the real
axis, and then the contour shrinks to $z=\pm 1$ from the outside of 
$C_+$ as $t\rightarrow \infty$. 
As the time decreases from $t=-1$, 
the contour becomes smaller with fixed boundaries and, at $t=-2$, the string
separates into two parts at the midpoint. As $t\rightarrow -\infty$, the
contour shrinks to $z=\pm 1$ from the inside of $C_+$.

Similarly, we can illustrate equal-time contours for propagation of
the right moving sector of a string by using
(\ref{eq:thetaminus}) and (\ref{eq:radius2}). As in
Fig.~\ref{fig:contour} (b), 
the resulting contours are given by rotating these
of the upper half plane by $180^\circ$
around the origin of the $z$ plane.

\begin{figure}
\begin{center}
\includegraphics[width=13cm]{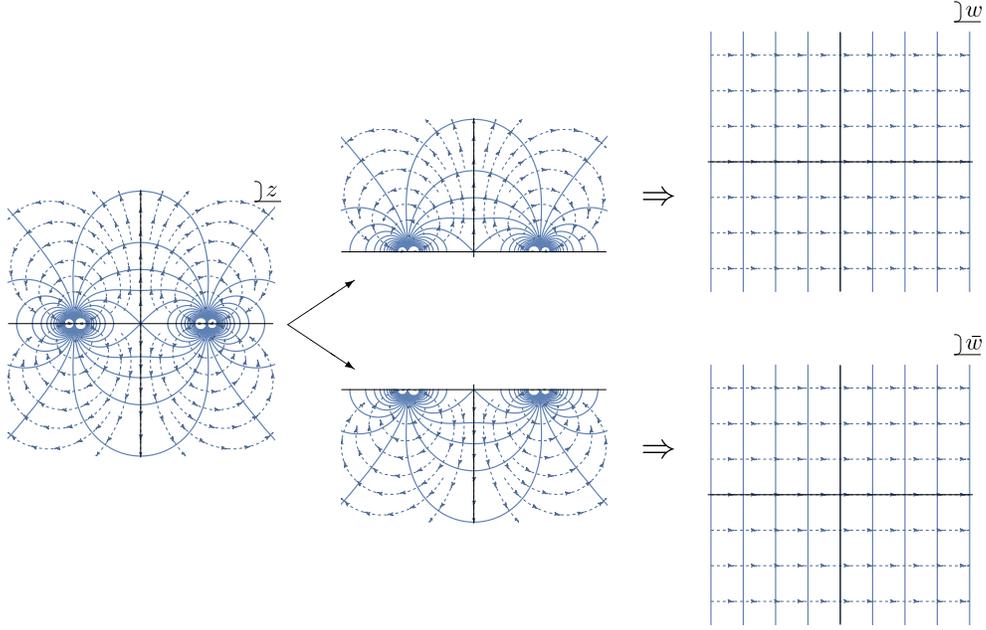}
\end{center}
\caption{
The equal-time contours generated by $H_g$ in the $z$ plane are divided into 
those of upper and lower half planes. 
 By introducing the complex
 coordinates, $w$ and $\bar{w}$, the $z$ plane is mapped to two complex
planes, which correspond to holomorphic and antiholomorphic parts for
 closed string worldsheets.}
\label{fig:map} 
\end{figure}

These contours have a remarkable feature that the string boundaries are
fixed at $z=\pm 1$ during propagation of the string. 
From (\ref{eq:ts}),  
one complex number $t+i s$ corresponds
to two points in the $z$ plane. Accordingly, we introduce a complex
coordinate $w=t+ is$ for the upper half $z$ plane and $\bar{w}=t+is$
for the lower half plane.
By this mapping, the upper half plane
corresponds to the whole $w$ plane, and the lower half plane to
the other $\bar{w}$ plane, as seen in Fig.~\ref{fig:map}.

Hence, the equal-time contours by
$H_g$ lead us to the worldsheet that consists of two complex planes.
The two planes, $w$ and $\bar{w}$, corresponding to the upper and lower half
$z$ planes are generated by the left and right moving Hamiltonians,
$H_g^+$ and $H_g^-$, respectively. Therefore, they can be
regarded as holomorphic and antiholomorphic worldsheets of a closed
string.

\subsection{Virasoro algebra for closed strings
\label{sec:virasoro}}

Now that we have obtained two decoupled Hamiltonians
for the left and right moving sectors,
we can construct
two independent Virasoro operators according to \cite{Ishibashi:2016bey}:
\begin{eqnarray}
 {\cal L}_\kappa= \int_{C_+^t}
\frac{dz}{2\pi i} g(z)f_\kappa(z) T(z),
~~~~
 \tilde{\cal L}_\kappa= \int_{C_-^t}
\frac{dz}{2\pi i} g(z)f_\kappa(z) T(z),
\end{eqnarray}
where $g(z)$ is the same function as that in the Hamiltonian (\ref{eq:Hg})
and $f_\kappa(z)$ is the function introduced in
\cite{Ishibashi:2016bey},
which is defined by the differential equation
\begin{eqnarray}
 g(z)\frac{\partial}{\partial z} f_\kappa(z)=\kappa f_\kappa(z).
\label{eq:deffk}
\end{eqnarray}
For a constant time $t$, $C_+^t$ and $C_-^t$ denote integral contours
along the equal-time line 
on the upper and lower half $z$ plane,
respectively. We should note again that $T(z)$ included in ${\cal
L}_\kappa$ and $\tilde{\cal L}_\kappa$ is the same energy-momentum
tensor of the open string system.

By setting $f_{\kappa=0}(z)=1$, ${\cal L}_0$ and $\tilde{\cal L}_0$
provide the left and right moving parts of the Hamiltonian (\ref{eq:Hg}),
i.e., ${\cal L}_0=H_g^+$ and $\tilde{\cal L}_0=H_g^-$. 
${\cal L}_\kappa$ satisfies the continuous Virasoro algebra
\begin{eqnarray}
 [{\cal L}_\kappa,\,{\cal L}_{\kappa'}]&=&
(\kappa-\kappa'){\cal L}_{\kappa+\kappa'}
\nn
&&+
\frac{c}{12}\int_{C_+^t}\frac{dz}{2\pi i}\left\{
(\kappa-\kappa')\left(\frac{\partial^2 g}{\partial z^2}
-\frac{1}{2g}\left(\frac{\partial g}{\partial z}\right)^2\right)
+\frac{\kappa^3-{\kappa'}^3}{2g}\right\}f_{\kappa+\kappa'}(z),
\label{eq:Virasoro}
\end{eqnarray}
which is antisymmetrized with respect to $\kappa$ and $\kappa'$ compared
to the equation given in \cite{Ishibashi:2016bey}.
Here, there is an important point in the
derivation of (\ref{eq:Virasoro}), which was not mentioned in
\cite{Ishibashi:2016bey}.
$g(z)$ in (\ref{eq:Hg}) has second- (or higher-) order zeros
at the boundaries of  $C_{\pm}^t$, $z=\pm 1$,
where $f_\kappa(z)$ for $\kappa\neq 0$ has essential singularities, 
which follow from the differential equation (\ref{eq:deffk}).
Accordingly, we need a regularization in addition to time-ordered
products for evaluation of ``equal-time'' commutators.

\begin{figure}
\begin{center}
\includegraphics[width=7.5cm]{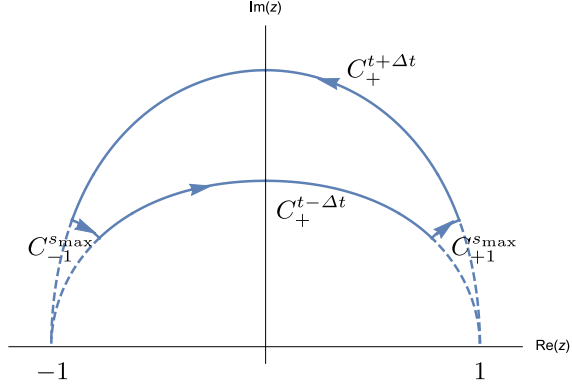}
\end{center}
\caption{Regularization for the essential singularities}
\label{fig:intcont}
\end{figure}

We illustrate this point in the simplest case of (\ref{eq:gsin2}).
By solving the differential equation (\ref{eq:deffk}), we find
\begin{eqnarray}
 f_\kappa(z)=\exp\left(\frac{2\kappa}{z^2-1}\right),
\end{eqnarray}
where we impose $f_0(z)=1$. $f_\kappa(z)$ has
essential singularities at $z=\pm 1$, which are 
second-order
zeros of $g(z)$.
In order to clarify the definition of ${\cal L}_\kappa$, we have
to avoid these singularities on the integration paths.
Here, we define ${\cal L}_\kappa$ by using an improper integral along
a string at the time $t$:
\begin{eqnarray}
 {\cal L}_\kappa=\lim_{s_{\rm max}\rightarrow \infty}
\int_{C_+^{t,s_{\rm max}}}
\frac{dz}{2\pi i} g(z)f_\kappa(z) T(z),
\label{eq:Ldef}
\end{eqnarray}
where $C_+^{t,s_{\rm max}}$ denotes the integration contour in which the
parameter region of $s$, which is introduced in (\ref{eq:ts}), is
limited to $-s_{\rm max}\leq s \leq s_{\rm max}$.
As usual, the ``equal-time'' commutator of ${\cal L}_\kappa$ and ${\cal
L}_{\kappa'}$ is regarded as a time-ordered product and so the
commutator is calculated as
\begin{eqnarray}
  [{\cal L}_\kappa,\,{\cal L}_{\kappa'}]&=&
 \lim_{
 \substack{
 \varDelta t\to +0\\
 s_{\rm max}\to +\infty
 }}
\int_{C_+^{t,s_{\rm max}}}\frac{dz'}{2\pi i}g(z')f_{\kappa'}(z')\times
\nn
&&\qquad\quad\times
\left(\int_{C^{t+\varDelta t,s_{\rm max}}_+}
-\int_{C^{t-\varDelta t,s_{\rm max}}_+}\right)
\frac{dz}{2\pi i}\,g(z)f_\kappa(z)\,
T(z)T(z').
\label{eq:intTT}
\end{eqnarray}
We notice that, unlike the radial quantization, it is impossible to
deform the integration paths of $z$ to a contour encircling $z'$ since the
paths have open boundaries associated with $\pm s_{\rm max}$.
To derive (\ref{eq:Virasoro}) by the path deformation,
we have to add the paths $C_{\pm 1}^{s_{\rm max}}$ as
depicted in Fig.~\ref{fig:intcont}, in which
$t$ varies from $t\mp \varDelta t$ to $t\pm \varDelta t$ 
with $s$ fixed to
$\mp s_{\rm max}$.\footnote{The equal-time contour in
Fig.~\ref{fig:intcont} represents the case in which the midpoint remains
joined on the $z$ plane.
Even in the case of splitting of the
midpoint, we only have to follow the
same procedure of adding $C_{\pm 1}^{s_{\rm max}}$.}
Then, what to be proved is that the integration along
$C_{\pm 1}^{s_{\rm max}}$ becomes zero as
$s_{\rm max}$ goes to infinity, i.e., as the paths shrink to the
essential singularity points. Fortunately, it can be found that the
limit of the integration is zero (see the discussion in
appendix~\ref{sec:intCpm}) and
the Virasoro algebra (\ref{eq:Virasoro}) holds for the operator defined
as (\ref{eq:Ldef}).

The right moving sector of the Virasoro operator $\tilde{\cal
L}_\kappa$ should also be defined by avoiding singularities along the
integration path on the lower half plane:
\begin{eqnarray}
 \tilde{\cal L}_\kappa=\lim_{s_{\rm max}\rightarrow \infty}
\int_{C_-^{t,s_{\rm max}}}
\frac{dz}{2\pi i} g(z)f_\kappa(z) T(z),
\label{eq:tildeLdef}
\end{eqnarray}
where $C_-^{t,s_{\rm max}}$ denotes the path in which the
parameter region of $s$ is
limited to $-s_{\rm max}\leq s \leq s_{\rm max}$.
Similarly, $\tilde{\cal L}_\kappa$ satisfies the
continuous Virasoro algebra.
Moreover, since
$C_+^{t,s_{\rm max}}$ and $C_-^{t,s_{\rm max}}$ have no intersections,
${\cal L}_\kappa$ and $\tilde{\cal L}_\kappa$ commute
with each other under this regularization:
\begin{eqnarray}
 [{\cal L}_\kappa,\,\tilde{\cal L}_{\kappa'}]=0.
\end{eqnarray}
Thus, we have found two independent Virasoro algebras in a deformed
open string system, which can be regarded as Virasoro algebras for
closed strings, i.e., the holomorphic and antiholomorphic parts.

\subsection{Natural frame for continuous Virasoro algebra}

The continuous Virasoro algebra (\ref{eq:Virasoro})
is explicitly calculated for the simplest weighting
function (\ref{eq:gsin2}):
\begin{eqnarray}
 \big[{\cal L}_\kappa,\,{\cal L}_{\kappa'}\big]&=&
 (\kappa-\kappa'){\cal L}_{\kappa+\kappa'}+
\frac{c}{12}\kappa^3\delta(\kappa+\kappa')
+\frac{3c}{96}(\kappa-\kappa')(1+|\kappa+\kappa'|)
e^{-(\kappa+\kappa'+|\kappa+\kappa'|)},
\label{eq:VirasoroEx1}
\\
 \big[\tilde{\cal L}_\kappa,\,\tilde{\cal L}_{\kappa'}\big]&=&
 (\kappa-\kappa')\tilde{\cal L}_{\kappa+\kappa'}+
\frac{c}{12}\kappa^3\delta(\kappa+\kappa')
+\frac{3c}{96}(\kappa-\kappa')(1+|\kappa+\kappa'|)
e^{-(\kappa+\kappa'+|\kappa+\kappa'|)}.
\label{eq:VirasoroEx2}
\end{eqnarray}
The term $\frac{c}{12}\kappa^3\delta(\kappa+\kappa')$ appears due to the
last term in (\ref{eq:Virasoro}) proportional to $\kappa^3$ and
this is also found in another example discussed in \cite{Ishibashi:2016bey}.
The third term in (\ref{eq:VirasoroEx1}) or (\ref{eq:VirasoroEx2})
is characteristic in this weighting function. But, 
it can be
absorbed in a shift of ${\cal L}_\kappa$,
since it is given as
$\kappa-\kappa'$ times a function of $\kappa+\kappa'$.

We consider this shift of ${\cal L}_\kappa$ in the context of conformal
transformation.
As discussed in \S \ref{sec:example}, we can
introduce a complex coordinate in the upper half plane, $w=t+is$,
which transforms the upper half $z$ plane to the whole $w$ plane.
Now, in the $w$ plane, we define a Virasoro operator associated with
${\cal L}_\kappa$:
\begin{eqnarray}
 \hat{\cal L}_\kappa=\int_{\hat{C}_+^t}\frac{dw}{2\pi i}e^{\kappa
  w}T(w),
\label{eq:wVirasoro}
\end{eqnarray}
where $\hat{C}_+^t$ is
an image of $C_+^t$ by the conformal transformation,
which is the path from $t-i\infty$ to $t+i\infty$ parallel
to the imaginary axis.
By the conformal transformation, $\hat{\cal L}_\kappa$ is related to
${\cal L}_\kappa$ as
\begin{eqnarray}
 \hat{\cal L}_\kappa={\cal
  L}_\kappa-\frac{c}{12}\int_{C_+^t}\frac{dz}{2\pi
  i}g(z)f_\kappa(z)\{w,\,z\},
\label{eq:hatL}
\end{eqnarray}
where $\{w,\,z\}$ is the Schwarzian derivative and, by using
$dz/dw=g(z)$, it is calculated as
\begin{eqnarray}
 \{w,\,z\}=
 \frac{1}{g}\left(
 \frac{1}{2g}\left(\frac{\partial g}{\partial z}\right)^2
-\frac{\partial^2 g}{\partial z^2}
\right).
\end{eqnarray}
We notice that this term is included in (\ref{eq:Virasoro}) as the term
proportional to $\kappa-\kappa'$
and so the shift (\ref{eq:hatL}) cancels this term in
the continuous Virasoro algebra. Consequently, we can find a simpler
form of continuous Virasoro algebra:
\begin{eqnarray}
 \big[\hat{\cal L}_\kappa,\,\hat{\cal L}_{\kappa'}\big]&=&
 (\kappa-\kappa')\hat{\cal L}_{\kappa+\kappa'}+
\frac{c}{12}\kappa^3\delta(\kappa+\kappa').
\label{eq:walgebra}
\end{eqnarray}
This algebra is the same as that in \cite{Ishibashi:2016bey}, 
though the
definition of the Virasoro operators and the weighting function $g(z)$
are different from those in \cite{Ishibashi:2016bey}.

Similarly, we can find antiholomorphic Virasoro algebra by
introducing a complex coordinate $\bar{w}$ in the lower half plane.
We define antiholomorphic
Virasoro operators for the path $\hat{C}_-^t$, the image of $C_-^t$:
\begin{eqnarray}
 {\hat{{\tilde{\cal L}}}}_\kappa=\int_{\hat{C}_-^t}
\frac{d\bar{w}}{2\pi i}e^{\kappa
  \bar{w}}T(\bar{w}),
\label{eq:wbarVirasoro}
\end{eqnarray}
and then we can find
\begin{eqnarray}
  \big[{\hat{\tilde{\cal L}}}_\kappa,\,\hat{\tilde{\cal
   L}}_{\kappa'}\big]
&=&
 (\kappa-\kappa'){\hat{\tilde{\cal L}}}_{\kappa+\kappa'}+
\frac{c}{12}\kappa^3\delta(\kappa+\kappa').
\label{eq:wbaralgebra}
\end{eqnarray}

Here, it should be noted that the definitions (\ref{eq:wVirasoro}) and
(\ref{eq:wbarVirasoro}) are independent of the choice of the weighting
function $g(z)$. As discussed above, 
for general $g(z)$ satisfying the conditions in  
\S \ref{sec:decoupling}, the deformed Hamiltonian is given by a sum of two
mutually commutative Hamiltonians, which are holomorphic and antiholomorphic
parts of a closed string Hamiltonian.
Then the Virasoro operators (\ref{eq:wVirasoro}) and
(\ref{eq:wbarVirasoro}) are defined through the natural coordinates
$w$ and $\bar{w}$ for the 2-sphere generated by the Hamiltonian.
Accordingly, the central term in (\ref{eq:Virasoro}) becomes irrelevant
to the choice of $g(z)$ due to
\begin{eqnarray}
 \int_{C_{\pm}^t} \frac{dz}{2\pi i}\frac{f_{\kappa+\kappa'}(z)}{g(z)}
=\int_{-\infty}^{\infty}\frac{ds}{2\pi}e^{(\kappa+\kappa')(t+is)}
=\delta(\kappa+\kappa').
\end{eqnarray}
Thus, the continuous Virasoro algebras (\ref{eq:walgebra}) and
(\ref{eq:wbaralgebra}) are universal for sine-square-like deformed
systems regardless of $g(z)$.

We can also derive the continuous Virasoro algebra
in the natural frame directly from the definitions
(\ref{eq:wVirasoro}) and (\ref{eq:wbarVirasoro}). To do so, we have only
to use the commutation relation (\ref{eq:TwTw'}) under the assumption that
the surface terms vanish at infinity. This assumption corresponds to the
fact that the integration along $C_{\pm1}^{s_{\rm max}}$ becomes zero in
the previous subsection.

\section{Closed string symmetries at the tachyon vacuum
\label{sec:CSS_TV}
}

\subsection{Identity-based tachyon vacuum solutions}

We consider cubic open bosonic string field theory. For the string field
$\Psi$ and the BRST operator $Q_{\rm B}$, the equation of motion is
given by $Q_{\rm B}\Psi+\Psi^2=0$.
In \cite{Takahashi:2002ez}, the equation was solved by using the
identity string field $I$ as
\begin{eqnarray}
 \Psi_0&=& Q_L(e^h-1)I-C_L((\partial h)^2 e^h)I,
\label{sol}
\end{eqnarray}
where $Q_L(f)$ and $C_L(f)$ are integrations of the BRST current
$j_{\rm B}(z)$ and the ghost $c(z)$,
which are multiplied by a function $f(z)$ along the
left half of a string. 
The function $h(z)$ satisfies $h(-1/z)=h(z)$ and $h(\pm i)=0$. Moreover,
the reality condition of (\ref{sol}) imposes 
$\left(h(z)\right)^*= h(1/z^*)$ on $h(z)$.

Expanding the string field around the solution $\Psi_0$, 
we obtain an action for fluctuation $\Psi$:
\begin{eqnarray}
S[\Psi;Q']=-\int \left(\frac{1}{2}
\Psi*Q'\Psi+\frac{1}{3}\Psi*\Psi*\Psi\right),
\label{eq:action}
\end{eqnarray}
where the modified BRST operator $Q'$ is given by
\begin{eqnarray}
 Q'&=& Q(e^h)-C((\partial h)^2 e^h).
\label{Qprime}
\end{eqnarray}
The operators $Q(f)$ and $C(f)$ are defined as integrations along a
whole unit circle.

The identity-based solution (\ref{sol}) includes an arbitrary function
$h(z)$ and it can be changed by gauge transformations. Since
the function continuously connects to zero,
most of the solutions are regarded as trivial pure gauge.
However, it is known that it provides the tachyon vacuum solution at the
boundary of some function spaces,
where $e^{h(z)}$ has second-order or higher-order zeros at some
points on the unit circle. If the second- (or higher-) order zeros exist, we
can construct a homotopy operator of (\ref{Qprime}) and it leads us to
the conclusion that we have no physical open string spectrum on the
identity-based tachyon vacuum \cite{Inatomi:2011xr}.
This vanishing cohomology, which was
solved earlier without the homotopy operator in \cite{Kishimoto:2002xi},
provides evidence 
that (\ref{sol}) correctly represents the tachyon vacuum solution.

For the moment, we assume that $e^{h(z)}$ has second-order zeros at $z=\pm
1$. We will discuss other cases later.

\subsection{Holomorphic and antiholomorphic decomposition of modified
BRST operators at the tachyon vacuum}

First, we decompose the modified BRST operator (\ref{Qprime}) into two
parts corresponding to the integrations along upper and lower
semicircles:
\begin{eqnarray}
 Q'=Q'_+ + Q'_-,
\end{eqnarray}
where $Q'_+$ and $Q'_-$ are defined by
\begin{eqnarray}
 Q'_{\pm}&\equiv&Q_{\pm}(e^h)-C_\pm((\partial h)^2 e^h),
 \label{eq:Q'pm_def}
\\
 Q_{\pm}(f)&\equiv&\int_{C_\pm}\frac{dz}{2\pi i}f(z)j_{\rm B}(z),
\label{eq:Qf}
\\
 C_{\pm}(f)&\equiv&\int_{C_\pm}\frac{dz}{2\pi i}f(z)c(z).
\label{eq:Cf}
\end{eqnarray}

Assuming that $f(z)$ and $g(z)$ have zeros at $z=\pm 1$, we can find the
following anti-commutation relations for $Q_{\pm}(f)$ and $C_{\pm}(f)$:
\begin{eqnarray}
 \{Q_{\pm}(f),\,Q_{\pm}(g)\}&=&
2\{Q_{\rm B},\,C_\pm(\partial f \partial g)\},
\label{eq:QpQp}
\\
  \{Q_{\pm}(f),\,C_{\pm}(g)\}&=&
\{Q_{\rm B},\,C_\pm(f g)\},
\label{eq:QpCp}
\\
 \{Q_{\pm}(f),\,Q_{\mp}(g)\}&=&
\{C_{\pm}(f),\,C_\mp(g)\}=
\{Q_{\pm}(f),\,C_\mp(g)\}=0.
\label{eq:QpQm}
\end{eqnarray}
A derivation is explained in appendix~\ref{sec:cr_modBRST}.
Alternatively, these relations are derived from the splitting algebra for
$Q_{L(R)}(f)$ and $C_{L(R)}(f)$ in \cite{Takahashi:2002ez},
which are defined by replacing the
integration paths $C_{\pm }$ with $C_{L(R)}$ in (\ref{eq:Qf}) and (\ref{eq:Cf}).
We have only to rotate the integration paths in the unit circle by
$90^\circ$ to obtain them because $C_{L(R)}$ is defined as right (left) unit semicircle. 

Now, we consider the identity-based tachyon vacuum solution generated by
$e^{h(z)}$ with second-order zeros only at $z=\pm 1$. In this case, owing to the
anti-commutation relations (\ref{eq:QpQp}), (\ref{eq:QpCp}) and
(\ref{eq:QpQm}), the operator $Q'$ at the tachyon vacuum splits
into two anti-commutative nilpotent operators (appendix~\ref{sec:cr_modBRST}):
\begin{eqnarray}
 (Q'_+)^2=(Q'_-)^2=\{Q_+',\,Q_-'\}=0.
\label{eq:nilpotent}
\end{eqnarray}
It should be noted that this decomposition of the modified BRST operator
occurs only for the tachyon vacuum solution. For trivial pure gauge
solutions, $e^{h(z)}$ has no zeros along the unit circle and so
(\ref{eq:nilpotent}) does not hold because the relations
(\ref{eq:QpQp}), (\ref{eq:QpCp}) and
(\ref{eq:QpQm}) cannot be realized for such a function due to surface
terms at $z=\pm 1$.
We notice that this decomposition is similar to that of the SSLD
Hamiltonian discussed in the previous section. By
analogy, $Q_+'$ and $Q_-'$ can be regarded as holomorphic and
antiholomorphic BRST operators of a closed string.

This interpretation is applicable to closed string states in the open
string field theory:
\begin{eqnarray}
 \ket{\cal V}_{\rm OSFT}=c(i)c(-i){\cal V}(i,-i)\ket{I},
\label{eq:calV}
\end{eqnarray}
where ${\cal V}(z,\bar{z})$ is a matter vertex operator for on-shell
closed string states and $\ket{I}$ is the ket representation
of the identity string field \cite{Zwiebach:1992bw, Hashimoto:2001sm, Gaiotto:2001ji}. 
It can be found that the state is
invariant separately for $Q_+'$ and $Q_-'$: 
\begin{eqnarray}
 Q_+'\ket{\cal V}_{\rm OSFT}= Q_-'\ket{\cal V}_{\rm OSFT}=0,
\label{eq:QpmV}
\end{eqnarray}
as in appendix \ref{sec:QprimeInv}.
On the other hand, in closed string theory, an on-shell closed string
state is given by
\begin{eqnarray}
 \ket{\cal V}=c(0)\tilde{c}(0){\cal V}(0,0)\ket{0},
\end{eqnarray}
where $\tilde{c}(\bar{z})$ is the antiholomorphic ghost field.
Since ${\cal V}$ is a $(1,1)$ primary operator, $\ket{\cal V}$ is also
invariant under the action of $Q_{\rm B}$,
the holomorphic part of the BRST operator,
 and 
the antiholomorphic counterpart $\tilde{Q}_{\rm B}$:
\begin{eqnarray}
 Q_{\rm B}\ket{\cal V}=\tilde{Q}_{\rm B}\ket{\cal V}=0.
\label{eq:QV}
\end{eqnarray}
These are analogous relations to (\ref{eq:QpmV}).
Correspondingly, $Q'_+$ and $Q'_-$ can be regarded as the BRST operators
$Q_{\rm B}$ and $\tilde{Q}_{\rm B}$ in closed string theories.

In closed string field theories, 
a gauge transformation for a closed string field $\Phi$
is given by
\begin{eqnarray}
 \delta\Phi =(Q_{\rm B}+\tilde{Q}_{\rm B})\Lambda+\cdots,
\end{eqnarray}
in which the summation of $Q_{\rm B}$ and $\tilde{Q}_{\rm B}$ is included.
Although an on-shell state is invariant separately as in (\ref{eq:QV}),
the quadratic term of the action for closed string field theory
is not invariant under the transformation
$\delta\Phi=Q_{\rm B}\Lambda +\tilde{Q}_{\rm B}\tilde{\Lambda}$.
Similarly, the open string field theory action (\ref{eq:action}) at the
tachyon vacuum is invariant under the gauge transformation
\begin{eqnarray}
 \delta\Psi=(Q'_+ + Q'_-)\Lambda +\Psi\ast\Lambda-\Lambda\ast\Psi,
\end{eqnarray}
but the quadratic term of (\ref{eq:action}) is not invariant for 
$\delta\Psi=Q'_+\Lambda_+ +Q'_-\Lambda_-$.
Thus, the open string field theory at the tachyon vacuum has an analogous
structure to closed string field theory under the correspondence of
the BRST operators.

\subsection{Energy-momentum tensor and Virasoro algebra}

On the perturbative vacuum, the string Hamiltonian in the Siegel gauge
is derived from the anti-commutation relation of the BRST operator with
the zero mode of 
the anti-ghost. 
Analogously, we can construct the Hamiltonian $H'$ at the tachyon vacuum
from $Q'_\pm$ and $b_0$:
\begin{eqnarray}
 H'&\equiv& \{Q',\,b_0\}= H_+ + H_-,
\\
H_\pm&=&\{Q'_{\pm},\,b_0\}
=
\int_{C_\pm}\frac{dz}{2\pi i} ze^{h(z)}\left(
T(z)+(\partial h)j_{\rm gh}(z)-(\partial h)^2(z)\right),
\label{eq:Htachyon}
\end{eqnarray}
where $T(z)$ is the total energy-momentum tensor given as $T(z)=\{Q_{\rm
B},\,b(z)\}$ and $j_{\rm gh}(z)$ is the ghost number current defined as
$j_{\rm gh}(z)=cb(z)$.
$H_+$ and $H_-$ commute with each other, because zeros of $e^{h(z)}$ are
now assumed to be second order at $z=\pm 1$ and then
the condition in (\ref{eq:Hg}) for SSLD, 
$g(\pm 1)=\partial g(\pm 1)=0$, is satisfied
(appendix~\ref{sec:cr_tv}). 
So, string propagations decompose to holomorphic and antiholomorphic
parts at the tachyon vacuum. Consequently, the identity-based tachyon
vacuum provides the SSLD system discussed in the previous section.

Let us consider finding the continuous Virasoro algebra at the tachyon
vacuum. It might be helpful to use the ghost twisted
energy-momentum tensor 
$T'(w)=T(w)-\partial j_{\rm gh}(w)$  ($z=e^w$) 
with the central charge $c=24$, since it is 
known 
that the Hamiltonian
$H'$ is expanded by modes of $T'(w)$ \cite{Takahashi:2003xe}.
However, 
continuous Virasoro algebra
using $T'(w)$ cannot be regarded as closed string
Virasoro algebra at the tachyon vacuum, because $T'(w)$ does not commute
with $Q'$.
Instead, we define an operator at the tachyon vacuum:
\begin{eqnarray}
 {\cal T}(z)&\equiv&e^{-h(z)}\{Q',\,b(z)\}
=T(z)+\partial h(z)\,j_{\rm gh}(z)-(\partial h(z))^2+\frac{3}{2}e^{-h(z)}
\partial^2 e^{h(z)}.
\label{eq:calT}
\end{eqnarray}
From the operator product expansions (OPEs) of $T(z)$ and $j_{\rm
gh}(z)$, we find that ${\cal T}(z)$ satisfies the same OPE as 
the energy-momentum tensor with zero central charge:
\begin{eqnarray}
 {\cal T}(y){\cal T}(z)\sim \frac{2}{(y-z)^2}{\cal T}(z)+\frac{1}{y-z}
\partial{\cal T}(z).
\end{eqnarray}
Here, it should be noted that ${\cal T}(z)$ includes not only operators
but also a function in its form. Since $h(z)$ is related to
a coordinate frame of worldsheets, ${\cal T}(z)$ has an explicit
dependence on the frame.
Admitting such an extension of operators, ${\cal T}(z)$ is interpreted as
the energy-momentum tensor at the tachyon vacuum.\footnote{The function
$h(z)$ is regarded as the operator that has regular OPEs with all
operators.}

By using ${\cal T}(z)$, we can define the continuous Virasoro operator at
the tachyon vacuum:
\begin{eqnarray}
 {\cal L}_\kappa \equiv \int_{C_+}\frac{dz}{2\pi i}
g(z)f_\kappa(z){\cal T}(z),
~~~~~
 \tilde{\cal L}_\kappa \equiv \int_{C_-}\frac{dz}{2\pi i}
g(z)f_\kappa(z){\cal T}(z),
\label{eq:calLkappa_def}
\end{eqnarray}
where the weighting function is related to $h(z)$ as $g(z)=z e^{h(z)}$.
Since $e^{h(z)}$ has second-order zeros at $z=\pm 1$, $g(z)$ also has 
second-order zeros at $z=\pm 1$.
As a result,
these operators satisfy the continuous Virasoro algebra for $c=0$
(\ref{eq:Virasoro}) and the antiholomorphic counterpart, and moreover
${\cal L}_\kappa$ and $\tilde{\cal L}_{\kappa'}$ commute with each other.
It can be easily found that ${\cal L}_0$ and $\tilde{\cal L}_0$ provide
holomorphic and antiholomorphic parts of the Hamiltonian
(\ref{eq:Htachyon}): ${\cal L}_0=H_+$ and $\tilde{\cal L}_0=H_-$.
By definition of ${\cal T}(z)$, these operators commute with
$Q^{\prime}_\pm$ (appendix~\ref{sec:BRST_tv}):
\begin{eqnarray}
 [ Q^{\prime}_\pm,\,{\cal L}_\kappa ] = [ Q^{\prime}_\pm,\,\tilde{\cal L}_\kappa ] =0.
\label{eq:3.23}
\end{eqnarray}
Thus, we have found the continuous Virasoro algebra in an open string
field theory at the tachyon vacuum.

\subsection{Ghost numbers}

As seen in (\ref{eq:calT}), the ghost sector of the energy-momentum
tensor is changed from the perturbative vacuum, while the matter sector is
unchanged. Accordingly, it is better to define ghost and antighost
fields at the tachyon vacuum as follows:
\begin{eqnarray}
 c'(z)\equiv e^{h(z)} c(z),~~~b'(z)\equiv e^{-h(z)} b(z).
\end{eqnarray}
These are also frame-dependent operators as ${\cal T}(z)$, but they
clearly satisfy $c'(y)b'(z)\sim 1/(y-z)$. Moreover, it can be easily
checked by the OPEs of ${\cal T}(z)$ 
that $c'(z)$ and $b'(z)$ are primary operators with
the conformal weights $-1$
and $2$, respectively. We note that, using this anti-ghost field,
(\ref{eq:calT}) is rewritten by a relation similar to that of the
perturbative vacuum:
\begin{eqnarray}
 {\cal T}(z)=\{Q',\,b'(z)\}.
\end{eqnarray}

We give a ghost number current at the tachyon vacuum by using $c'(z)$
and $b'(z)$:
\begin{eqnarray}
 {\cal J}_{\rm gh}(z)=c'\,b'(z).
\end{eqnarray}
Since it should be defined by the normal-ordering prescription
\begin{eqnarray}
 {\cal J}_{\rm gh}(z)=\lim_{y\rightarrow z}
\left[c'(y)b'(z)-\frac{1}{y-z}\right],
\end{eqnarray}
the current ${\cal J}_{\rm gh}(z)$ is related to the conventional
ghost number current as
\begin{eqnarray}
 {\cal J}_{\rm gh}(z)=j_{\rm gh}(z)+\partial h(z).
\label{eq:calJ}
\end{eqnarray}
This is also a frame-dependent operator due to $h(z)$ and we can easily find that it
satisfies the same OPEs as those of the perturbative vacuum:
\begin{eqnarray}
 {\cal T}(y){\cal J}_{\rm gh}(z)&\sim&
  \frac{-3}{(y-z)^3}+\frac{1}{(y-z)^2}{\cal J}_{\rm gh}(z)
+\frac{1}{y-z}\partial {\cal J}_{\rm gh}(z),
\\
{\cal J}_{\rm gh}(y){\cal J}_{\rm gh}(z)
&\sim& \frac{1}{(y-z)^2}.
\end{eqnarray}

Now that a ghost number current is given at the tachyon vacuum, we can
define
an operator counting the ghost number with respect to $c'(z)$
and $b'(z)$:
\begin{eqnarray}
 {Q_{\rm c}'}_\pm \equiv 
\int_{C_\pm}\frac{dz}{2\pi i}\,{\cal J}_{\rm gh}(z).
\end{eqnarray}
These satisfy the following commutation relations with the BRST
operators (appendix~\ref{sec:BRST_tv}):
\begin{eqnarray}
 [{Q'_{\rm c}}_\pm,\,Q'_\pm]=Q'_\pm,
~~~~
 [{Q'_{\rm c}}_\pm,\,Q'_\mp]=0.
 \label{eq:3.32}
\end{eqnarray}
According to the correspondence between $Q'_\pm$ and the closed string
BRST operators, these imply that ${Q'_{\rm c}}_+$ and ${Q'_{\rm c}}_-$
count 
the holomorphic and antiholomorphic ghost numbers, respectively.

Here, it is interesting to consider the relation of ${Q'_{\rm c}}_\pm$
to the conventional ghost number $Q_{\rm c}$. From the relation
(\ref{eq:calJ}), we have
\begin{eqnarray}
 {Q'_{\rm c}}_+ + {Q'_{\rm c}}_-=Q_{\rm c}+\oint_{C_{+}+C_{-}}\frac{dz}{2\pi
  i}\partial h(z).
\end{eqnarray}
So, this might suggest that the ghost number for open strings has a linear
relation to the ghost number for closed strings.
However, the integration on the right-hand side is ill-defined due to
poles on the unit circle, which arise from the relation $\partial
h(z)=(\partial e^{h(z)})/e^{h(z)}$ and the second-order zeros of
$e^{h(z)}$ at $z=\pm 1$. Consequently, we have no linear relations
between ${Q'_{\rm c}}_\pm$ and $Q_{\rm c}$. This result reflects the
fact that the ghost number for open strings is not definable at the
tachyon vacuum since we have no open strings there.
In other words, open string states belong to a completely different
space to that of closed string states.

\subsection{Frame-dependent operators and similarity transformations}

So far, we have found ${\cal T}(z)$, $c'(z)$, $b'(z)$ and
${\cal J}_{\rm gh}(z)$, which satisfy the same OPEs as those of the
conventional operators,  although they depend on the frame in the
definition.
These operators can be better understood by a similarity
transformation generated by
\begin{eqnarray}
 q(h)=\oint \frac{dz}{2\pi i}h(z)j_{\rm gh}(z).
\end{eqnarray}
A transformation of this type was first used to prove vanishing
physical cohomology at the tachyon vacuum \cite{Kishimoto:2002xi}.
As in \cite{Kishimoto:2002xi}, we can obtain the relation
\begin{eqnarray}
 Q'=e^{q(h)}Q_{\rm B}e^{-q(h)},
\end{eqnarray}
but this is rather formal on the tachyon vacuum, because $e^{q(h)}$
becomes divergent if it is represented in the normal ordered form
 in the conventional Fock space.
However, this formal transformation is useful in understanding the
operators at the tachyon vacuum.
We find that all of the operators can be rewritten as the formal
transformation of the conventional operators:
\begin{eqnarray}
&&
c'(z)=e^{q(h)}c(z)e^{-q(h)},
~~~~
b'(z)=e^{q(h)}b(z)e^{-q(h)},
\\
&&
{\cal T}(z)=e^{q(h)}T(z)e^{-q(h)},
~~~~
{\cal J}_{\rm gh}(z)=e^{q(h)}j_{\rm gh}(z)e^{-q(h)}.
\end{eqnarray}
These directly lead to the same OPEs and commutation relations as those
of the perturbative vacuum.
Moreover,
the similarity transformation is given as a Bogoliubov-type transformation
and the inner product of the vacua related by the transformation 
is ill-defined because $e^{q(h)}$ becomes divergent in the original Fock space.
Therefore, the ghost number at the tachyon vacuum can never be
related to that of the perturbative vacuum as discussed above.

We can apply the notion of frame-dependent operators to the BRST current.
By the similarity transformation generated by $q(h)$,
the conventional BRST current $j_{\rm B}(z)$ is transformed to
\begin{eqnarray}
 {\cal J}_{\rm B}(z)= e^{h(z)}j_{\rm B}(z)
+\{2\partial^2 h(z)+(\partial
  h)^2(z)\} e^{h(z)} c(z)
+2\partial h(z) e^{h(z)} \partial c(z).
\label{eq:3.38}
\end{eqnarray}
As well as other operators, ${\cal J}_{\rm B}(z)$ satisfies the same OPEs
as the conventional ones. In addition, the modified BRST operator
$Q'=Q_+'+Q_-'$ can
be rewritten by using ${\cal J}_{\rm B}(z)$ 
(\ref{eq:app_calJB-jB'}):
\begin{eqnarray}
 Q'_\pm =\int_{C_\pm} \frac{dz}{2\pi i} {\cal J}_{\rm B}(z).
 \label{eq:3.39}
\end{eqnarray}

\subsection{Other types of identity-based tachyon vacuum solutions}

We have seen that the identity-based tachyon vacuum solution
leads to closed string symmetries in open SFT by using results of SSLD,
provided that $e^{h(z)}$ has second-order zeros at $z=\pm 1$.
In this subsection, we discuss other types of 
identity-based vacuum solutions.

First, we consider the solution for $e^{h(z)}$ with higher-order zeros at
$z=\pm 1$, which must be even order due to the hermiticity condition for
$h(z)$.
It is known that this type of solution provides the same vacuum
structure for 
the expanded theory \cite{Igarashi:2005wh} and a homotopy operator
for the modified BRST operator \cite{Inatomi:2011xr}.
Accordingly, it is also regarded as
a tachyon vacuum solution.
In this case, the weighting function $g(z)=z e^{h(z)}$ has higher-order 
zeros at $z=\pm 1$.
Since they satisfy $g(\pm 1)=\partial g(\pm 1)=0$, decomposition
of the modified BRST operator occurs and so we find holomorphic and
antiholomorphic continuous Virasoro algebra.
Similarly to the case of second-order zeros, we can find closed string
symmetries on this vacuum.

\begin{figure}
\begin{center}
\includegraphics[width=16cm]{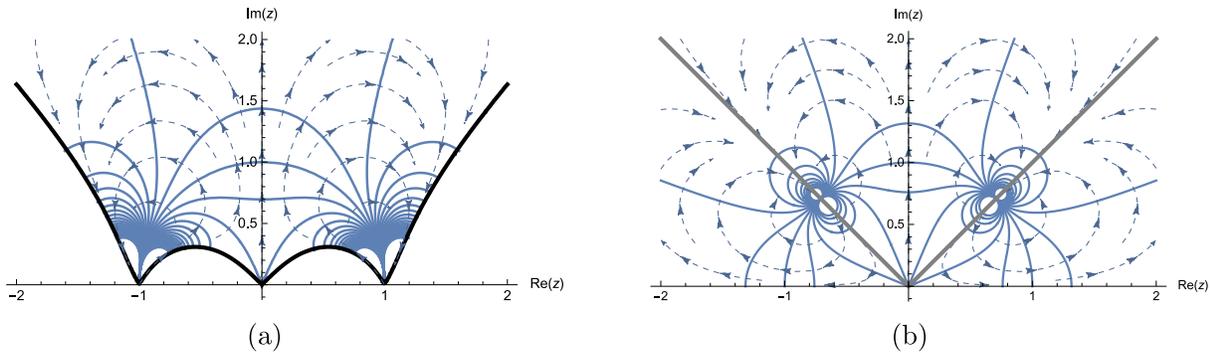}
\end{center}
\caption{The equal-time contours of a string for: (a) the weighting
 function (\ref{eq:4thzerofunc}), which has forth-order zeros at $z=\pm
 1$; and (b)\ the weighting function (\ref{eq:gzZ4}), which has zeros
 except at $z=\pm 1$. 
 (Dashed lines with arrows denote evolution of time $t$.
 Thick lines denote branch cuts.)}
\label{fig:contour2} 
\end{figure}

Here, we consider the difference 
between these two cases. For example, we
consider the following function with
the fourth-order zeros at $z=\pm 1$ \cite{Igarashi:2005wh}:
\begin{eqnarray}
 g(z)
 =ze^{h(z)}= \frac{z}{16}\left(z-\frac{1}{z}\right)^4.
\label{eq:4thzerofunc}
\end{eqnarray}
As discussed in \S\ref{sec:example}, we can introduce the
complex
coordinates $w$ and $\bar{w}$ on the worldsheet generated by ${\cal
L}_0$ and $\tilde{\cal L}_0$.
$w=t+is$ is given by the integral of $1/g(z)$:
\begin{eqnarray}
 t+is=\frac{4(1-3z^2)}{3(z^2-1)^3}.
\end{eqnarray}
The equal-time contour for the holomorphic sector is depicted
in Fig.~\ref{fig:contour2} (a).
The contour at $t=-2/3$ is given by the half unit circle.
The half unit circle becomes larger as time goes, but open string
boundaries are fixed at $z=\pm 1$. At $t=0$, the string midpoint goes
to infinity and then the string apparently splits into two parts, but
the left and right cuts, which are the thick curve in Fig.~\ref{fig:contour2}
(a),
are identified to each other. We note that the curve is given
by the equation $s=0$. Similarly, as time
goes back, the half unit circle becomes smaller and before the time $t=-4/3$
the midpoint moves on the branch curve.
As $t\rightarrow \pm \infty$,
the contour shrinks to the points $z=\pm 1$.

As seen in Fig.~\ref{fig:contour2} (a), a string propagates in
a partial region of the $z$ plane, and this is a different point from
the previous case in Fig.~\ref{fig:contour} (a).
This behavior is clarified by the differential equation
\begin{eqnarray}
 \frac{dw}{dz}=\frac{1}{g(z)}.
\end{eqnarray}
If $g(z)$ has zeros as $g(z)\sim a_\pm (z\mp 1)^n+\cdots$,
$w$ is expanded near the zeros as
\begin{eqnarray}
 w\sim - \frac{1}{(n-1)a_\pm}\, (z\mp 1)^{-n+1}+\cdots.
\end{eqnarray}
From this asymptotics, we find that $w$ goes around infinity if $z$
rotates by $2\pi/(n-1)$ around $z=\pm 1$. As a result,
the $z$ plane is divided into $n-1$ parts due to the branch cuts.
Therefore, in the second-order
case ($n=2$), the region swept by a string
consists of one piece of the whole $z$ plane, and in the forth-order
case ($n=4$), it is a partial region of the $z$ plane.
We note that the angle of the region at $z=\pm 1$ is $\pi/(n-1)$.

Secondly, let us consider the solution generated by
\begin{eqnarray}
 e^{h(z)}=\frac{(-1)^l}{4}\left(z^l + \frac{(-1)^l}{z^l}\right)^2
\ \ \ (l=1,2,\dots).
\end{eqnarray}
The case of $l=1$ corresponding to (\ref{eq:gsin2})
 is discussed in the previous section as the second-order-zero case. 
 These solutions were studied in \cite{Kishimoto:2002xi, Igarashi:2005sd, Inatomi:2011xr} and
vanishing physical cohomology was proved. All the numerical studies and
homotopy operators indicate that these are the same tachyon vacuum solution
as the case of $l=1$ \cite{Takahashi:2003ppa,Kishimoto:2009nd,Kishimoto:2011zza}.
However, these solutions have zeros at points 
other than $z=\pm 1$
and, particularly for the even-$l$ cases, they do not
have zeros at $z=\pm 1$. So we cannot apply the SSLD mechanism in
the previous section directly to these cases.  

As a characteristic example, we consider the weighting function for $l=2$:
\begin{eqnarray}
 g(z)=\frac{z}{4}\left(z^2+\frac{1}{z^2}\right)^2.
\label{eq:gzZ4}
\end{eqnarray}
This function has zeros at the fourth roots of $-1$,
$z=\exp(\pm i(2k-1)\pi/4)\ \ (k=1,2)$,
which are not equal to $\pm 1$.
The complex coordinate for the holomorphic part
is given by
\begin{eqnarray}
 w=\frac{-1}{1+z^4},
\end{eqnarray}
and then the unit semicircle ($t={\rm Re}\,w=-1/2$) 
moves as depicted
in Fig.~\ref{fig:contour2} (b). From this figure, we notice that
the open string boundaries, which are on the real axis, move while they
are static at $z=\pm 1$ in the case of $l=1$ (Fig.~\ref{fig:contour} (a)),
but the zeros  
of (\ref{eq:gzZ4})
are fixed during the propagation of a string in the $z$ plane.

\begin{figure}
\begin{center}
\includegraphics[width=14.5cm]{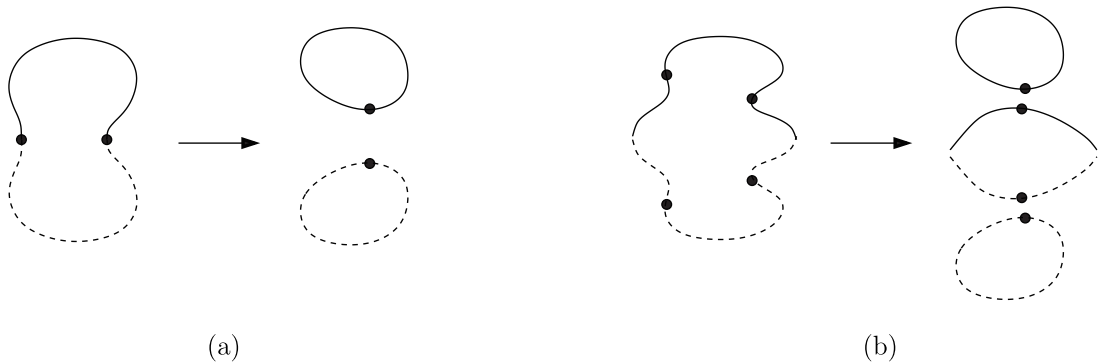}
\end{center}
\caption{String pictures before and after SSLD. The solid and dashed lines
 correspond to holomorphic and antiholomorphic parts of a string.
(a)
 The case of the weighting function (\ref{eq:gsin2}). As a result of
 SSLD, open string boundaries (black dots) become joined and an open
 string divides to holomorphic and antiholomorphic strings.
(b) The
 case of (\ref{eq:gzZ4}). The black dots represent the points of zeros
 of the weighting function. SSLD leads to the division of an open string
 into one closed and one open string.}
\label{fig:ssd_string} 
\end{figure}

Before considering the $z$ plane, we recapitulate the
SSLD mechanism in terms of
the string picture. In SSLD, open string boundaries no longer propagate
due to zeros of $g(z)$ and so the worldsheet generated by such a string
becomes a closed surface without boundaries. From the viewpoint of
the doubling trick,
a sphere consisting of holomorphic and antiholomorphic disks is pinched
along an open string boundary, and then it splits into two spheres,
which correspond to holomorphic and antiholomorphic surfaces of a
closed string. Operators on the holomorphic sphere are uncorrelated with
those on the antiholomorphic sphere due to commuativity of
the Virasoro operators, $[{\cal L}_\kappa,\,\tilde{\cal L}_{\kappa'}]=0$.
After all, the pinch of an open string boundary means that the endpoints
of an open string are bounded each other and then the open string
becomes a closed string in SSLD. In other words, a closed string as a
result of the doubling trick is pinched at two points corresponding
to the open string endpoints and then it separates to two closed strings,
as depicted in Fig.~\ref{fig:ssd_string} (a).

We apply this picture of SSLD to 
the case of $l=2$.
If we use the usual parametrization of an open string with
$\sigma\ \ (0\leq \sigma\leq \pi)$, the points $\sigma=\pi/4,\ 3\pi/4$
do not propagate in this vacuum due to the zeros of $g(z)$.
The open string is bounded at these points and it separates to a closed
string, which corresponds to $\pi/4<\sigma<3\pi/4$,
and an open string as in Fig.~\ref{fig:ssd_string} (b).
Finally, we obtain an open string in addition to a closed string
at this vacuum, although we have no open strings at the tachyon vacuum.

This inconsistency is resolved by considering physical observables in open
SFT. A gauge invariant observable in open SFT is given by
\begin{eqnarray}
  \big<\Psi\big|{\cal V}\big>_{\rm OSFT}\,,
\end{eqnarray}
where $\Psi$ is an open string field and $\big|{\cal V}\big>_{\rm OSFT}$
defined in (\ref{eq:calV}) corresponds to an on-shell
closed string vertex operator. The point is that the closed string vertex operator
in $\big|{\cal V}\big>_{\rm OSFT}$ is inserted at $\sigma=\pi/2$.
Therefore, at this vacuum, only the resulting closed string can be coupled
with the closed string vertex operator and the separated open string has
no interaction with the vertex. If we calculate perturbatively a
correlation function of these observables, the resulting closed string
is coupled to the external closed string vertex and 
so provides closed string worldsheets
with the vertex. However, the separated open string gives worldsheets with
no external vertex, i.2., stringy vacuum bubble diagrams.
Hence, the closed string consisting of the inner part of an open string
contributes to physical observables, but the open string consisting of
the outer part of an open string is completely irrelevant to the closed
vertex operator.

Returning to the $z$ plane of string propagation, we have branch cuts along
$\arg z=\pi/4,3\pi/4$
as a result of the decomposition of an open string at this vacuum.
The upper quarter plane includes the trajectory of the string midpoint
$\sigma=\pi/2$.
Since the branch cuts 
$\arg z=\pi/4$ and $\arg z=3\pi/4$
are identified, the upper quarter plane provides a holomorphic closed
string surface 
as in the case of $l=1$.
As a result of SSLD, only
the correlation function on this plane contributes to observables.
As far as physical observables are concerned, we conclude that the
weighting function with 
(\ref{eq:gzZ4})
also provides holomorphic and antiholomorphic sectors of a
closed string.

\section{Concluding remarks
\label{sec:Conclusion}
}

In this paper, we have studied SSLD in open string systems and clarified
that the left and right moving modes in the SSLD system are decoupled
and uncorrelated by zeros of the weighting function of the Hamiltonian.
We have shown that, as a result of the decoupling, the SSLD system is
equivalent to a closed string system, in the sense that we find 
holomorphic and antiholomorphic Virasoro algebra in the SSLD system.

Next, we considered open SFT expanded around the identity-based
tachyon vacuum solution. We have found that the modified BRST operator
is decomposed to holomorphic and antiholomorphic parts,
which are anti-commutative and nilpotent. 
These operators are analogous to
closed string BRST operators and so the gauge
symmetry of the theory is regarded as that of closed string field
theories. 
By using these BRST operators, we have constructed
the local operators in the tachyon vacuum, including
the energy-momentum tensor, ghost and anti-ghost fields, 
the ghost number current, and the BRST current.
It is a remarkable feature that these operators depend on the frame of
the worldsheet, which is chosen by the identity-based tachyon vacuum
solution. From these operators, we have found the continuous
Virasoro algebra and the ghost number operator. The important point is
that these have holomorphic and antiholomorphic parts, which are
realized by the SSLD mechanism.

The theory at the identity-based tachyon vacuum solution possesses a gauge
symmetry generated by the holomorphic and antiholomorphic BRST
operators, which is identified with a gauge symmetry of closed string
field theories. Since gauge symmetry is an essential ingredient in SFT,
we conjecture that the theory at the tachyon vacuum provides a kind of
closed string field theory.
That is, it is expected that the theory includes the dynamics of closed
strings, although the field in the theory is an open string field.
Actually, observables of pure closed
strings might be calculable in terms of gauge invariants in this theory.
It is known that a similar situation has occurred in matrix theories,
where closed string
amplitudes are obtained in terms of matrices replacing an open string field
\cite{Banks:1996vh,Ishibashi:1996xs}.
More interestingly, the theory has only a cubic interaction, which is
simpler than conventional covariant closed string field theories.
As a first step to such a closed string field theory,
it will be worth finding a 
method to calculate the observables of closed strings in this vacuum.

We should comment on the cohomology of the modified BRST operator at the
tachyon vacuum. As proved in \cite{Kishimoto:2002xi,Inatomi:2011xr}, the
modified BRST operator has vanishing cohomology and so the theory seems
to have not only no open strings but also no closed strings at the
tachyon vacuum.
This apparent contradiction is immediately resolved if one considers
closed strings at the perturbative vacuum. The BRST cohomology at
the perturbative vacuum includes a physical open string spectrum only, but
on-shell closed string states can be introduced
 as the state (\ref{eq:calV}),
which provides a gauge invariant observable in SFT.
Therefore, at the perturbative vacuum, a correlator of the gauge
invariant observables leads to a closed string
amplitude \cite{Zwiebach:1992bw}, which is
explicitly calculated in \cite{Takahashi:2003kq}.
Similarly, a closed string state is
included as the gauge invariant observable given by the state
(\ref{eq:calV}) even at the tachyon vacuum
and so there is a possibility of deriving closed string observables
in spite of the vanishing cohomology.

While closed string states are related to the state (\ref{eq:calV}),
it is important to clarify the representation of the continuous Virasoro
algebra. 
It is known that the vacuum of the discrete SSD
system is exactly equal to that of the closed
system \cite{KatsuraEGS}.
Analogously, the vacuum for the continuous Virasoro algebra is expected
to correspond to that of closed string theories,
although the correspondence is still unclear.
Furthermore, it is much more difficult to understand excited states for the continuous Virasoro algebra.
It will be interesting in future work to investigate representations of the
continuous Virasoro algebra, if possible, by referring to SFT. 
In an appropriate representation space, the cohomology of the modified BRST operator
might be nontrivial and related to closed string states.

Finally, we should comment on the level-matching condition, which is
imposed on a conventional closed string field to assure the $\sigma$
translational invariance of a closed string. In conventional closed
SFTs, 
it is imposed on a closed
string field as $(L_0-\tilde{L}_0)\Phi=0$. However, it is difficult to
find 
such a condition in open SFT on the
identity-based tachyon vacuum. For instance, it is easily found that the
open string field or the closed string state do not satisfy a 
straightforward
extension of the level-matching condition: $({\cal L}_0-\tilde{\cal
L}_0)\Psi=0$ or $({\cal L}_0-\tilde{\cal L}_0)\ket{\cal V}_{\rm
OSFT}=0$. This is not a problem because the open string field and
the closed string state 
do not have
uniformity due to the fixed boundary
points or the midpoint of a string. This suggests that the level-matching
condition is 
not
necessarily required in closed string field theories and
our theory indicates the possibility of a closed SFT without the condition.
On the other hand, this possibility is pursued in SFT in terms of a
conventional closed string field \cite{SFT2018okawa}.
It will be interesting to understand
the relation between these SFTs.

\section*{Acknowledgments}
The authors are grateful to Nobuyuki Ishibashi and Tsukasa Tada for valuable
discussions and comments.
We would like to thank Mayuri Kasahara for her help at
the beginning of this work. 
T.~T. would also like to thank the
organizers of the workshop on 
``Sine square deformation and related topics''
at RIKEN,
where this work was initiated, for their hospitality.
T.~K. would also like to thank the Yukawa Institute for Theoretical Physics at Kyoto
University for its hospitality during the workshop ``Strings and Fields
2018''.
This work was supported in part by a JSPS Grant-in-Aid for Scientific
Research (C) (JP15K05056).

\appendix

\section{Commutation relations for the SSLD Hamiltonian
\label{sec:cr_SSLD}}

For the energy-momentum tensor on the $z$ plane,
$
 T(z)=\sum_{n\in\mathbb{Z}}L_nz^{-n-2}
$,
satisfying the OPE
\begin{align}
T(z)T(z^{\prime})\sim \frac{c/2}{(z-z^{\prime})^4}+\frac{2}{(z-z^{\prime})^2}T(z^{\prime})
+\frac{1}{z-z^{\prime}}\partial T(z^\prime),
\end{align} 
we have the Virasoro algebra with the central charge $c$ :
\begin{align}
 [L_m,L_n] = (m-n)L_{m+n} +\frac{c}{12}(m^3-m)\delta_{m+n,0}.
\end{align}
With this algebra, we can derive (\ref{eq:TT})  as
\begin{align}
 [T(z),T(z^{\prime})]&=\sum_{m,n\in \mathbb{Z}}z^{-m-2}z^{\prime -n-2}
 \left(
  (m-n)L_{m+n} +\frac{c}{12}(m^3-m)\delta_{m+n,0}
 \right)\nonumber\\
 &=\sum_{m,l\in \mathbb{Z}}(m+1)z^{-m-2}z^{\prime m-l-2}L_l
 -\sum_{n,l\in\mathbb{Z}}(n+1)z^{n-l-2}z^{\prime -n-2}L_l
 -\frac{c}{12}\partial^3\sum_{m\in\mathbb{Z}}z^{-m+1}z^{\prime m-2}
 \nonumber\\
 &=-\partial\sum_{m\in \mathbb{Z}}z^{-m-1}z^{\prime m}T(z^{\prime}) 
 +\partial^{\prime}\sum_{n\in\mathbb{Z}}z^nz^{\prime -n-1}T(z)
 -\frac{c}{12}\partial^3\sum_{m\in\mathbb{Z}}z^{-m+1}z^{\prime m-2}
 \nonumber\\
 &=-\partial\delta(z,z^{\prime})T(z^{\prime})+\partial^{\prime}\delta(z,z^{\prime})T(z)
 -\frac{c}{12}\partial^3\delta(z,z^{\prime})
 \nonumber\\
&=-(T(z)+T(z^{\prime}))\partial\delta(z,z^{\prime})
-\frac{c}{12}\partial^3\delta(z,z^{\prime}),
\label{eq:TT_proof}
\end{align}
($\partial=\frac{\partial}{\partial z},~\partial^{\prime}=\frac{\partial}{\partial z^{\prime}}$), 
where we have used the notion that
the delta function on the unit circle, $\delta(z,z^{\prime})
=\sum_{n\in \mathbb{Z}}z^{-n-1}z^{\prime n}$, 
satisfies
\begin{align}
\partial\delta(z,z^{\prime})&=-\partial^{\prime}\delta(z,z^{\prime}).
\label{eq:partial_delta}
\end{align}
Actually, the commutation relation (\ref{eq:TT}) holds for any frame. 
For a conformal transformation $w=w(z)$, the energy-momentum tensor $T(w)$
is given by
$T(w)=(\frac{dz}{dw})^2\left(T(z)-\frac{c}{12}\{w,z\}\right)$,
where 
$\{w(z),\,z\}=\frac{w^{\prime\prime\prime}}{w^{\prime}}-\frac{3}{2}
(\frac{w^{\prime\prime}}{w^{\prime}})^2$
 is the Schwarzian derivative,
 and the commutation relation is computed with  (\ref{eq:TT})
 as follows:
\begin{align}
  [T(w),T(w^{\prime})]&=\left(\frac{dz}{dw}\right)^2\left(\frac{dz^{\prime}}{dw^{\prime}}\right)^2
  \left(-(T(z)+T(z^{\prime}))\partial\delta(z,z^{\prime})
-\frac{c}{12}\partial^3\delta(z,z^{\prime})\right)
\nonumber\\
&=-\left[T(w)\left(\frac{dz^{\prime}}{dw^{\prime}}\right)^2+T(w^{\prime})\left(\frac{dz}{dw}\right)^2
\right]\partial\delta(z,z^{\prime})
\nonumber\\
&\quad\, -\frac{c}{12}\left(\frac{dz}{dw}\right)^2\left(\frac{dz^{\prime}}{dw^{\prime}}\right)^2
\left[\partial^3\delta(z,\,z')
+\left(\{w,\,z\}+\{w',\,z'\}
\right)\partial\delta(z,\,z')
\right].
\label{eq:TwTw'_temp}
\end{align}
Here, for the transformation $w=w(z)$, 
we define the delta function as
\begin{eqnarray}
 \delta(w,\,w')=\frac{dz}{dw}\delta(z,\,z')=\frac{dz'}{dw'}\delta(z,\,z').
\end{eqnarray}
The integration range is changed to
a contour $C$ for $w$, which is the image of the unit circle on the $z$ plane.
Corresponding to (\ref{eq:delta_normalization}), its normalization is given by
\begin{eqnarray}
 \int_C \frac{dw}{2\pi i} \delta(w,\,w')=1.
\end{eqnarray}
For the delta function, we find the following relations:
\begin{eqnarray}
&&
 \partial_w \delta(w,\,w')=\frac{dz}{dw}\frac{dz'}{dw'}
\partial \delta(z,\,z')
=\left(\frac{dz}{dw}\right)^2\partial\delta(z,z^{\prime})+\frac{d^2z}{dw^2}\delta(z,z^{\prime})
=-\partial_w^{\prime}\delta(w,w^{\prime}),
\label{eq:d1delta_w}
\\
&&
\partial_w^3 \delta(w,\,w')
=\left(\frac{dz}{dw}\right)^2\left(\frac{dz'}{dw'}\right)^2
\left[\partial^3\delta(z,\,z')
+\left(\{w,\,z\}+\{w',\,z'\}
\right)\partial\delta(z,\,z')
\right],
\label{eq:d3delta}
\end{eqnarray}
where $\partial_w=\frac{\partial}{\partial w},~\partial_w^{\prime}=\frac{\partial}{\partial w^{\prime}}$.
By using (\ref{eq:d1delta_w}) and (\ref{eq:d3delta}), 
(\ref{eq:TwTw'_temp}) can be rewritten as:
\begin{eqnarray}
 [T(w),\,T(w')]=-(\,T(w)+T(w')\,)\,\partial_w
  \delta(w,w')-\frac{c}{12}\partial_w^3 \delta(w,w'),
  \label{eq:TwTw'}
\end{eqnarray}
which is the same form as (\ref{eq:TT}).

For the left and right moving modes of the Hamiltonian, $H^{\pm}_g$ (\ref{eq:Hg}),
the commutation relation (\ref{eq:TT}) is applied as
\begin{align}
 [H^+_g,H^-_g]&=\int_{C_+}\frac{dz}{2\pi i}\int_{C_-}\frac {dz^{\prime}}{2\pi i}
 g(z)g(z^{\prime})\left(
 -(T(z)+T(z^{\prime}))\partial\delta(z,z^{\prime})
-\frac{c}{12}\partial^3\delta(z,z^{\prime})
 \right)
 \nonumber\\
 &=\int_{C_+}\frac{dz}{2\pi i}\int_{C_-}\frac {dz^{\prime}}{2\pi i}
\left( g(z^{\prime})\left(\partial g(z)(T(z)+T(z^{\prime}))
 +g(z)\partial T(z)\right)
-\frac{c}{12}\partial^2 g(z)\partial g(z^{\prime})
 \right)\delta(z,z^{\prime})
 \nonumber\\
&=0,
\label{eq:app_Hg+Hg-=0}
\end{align}
where we have used the condition 
$g(\pm 1)=\partial g(\pm 1)=0$, eq.~(\ref{eq:partial_delta}) and
\begin{align}
\int_{C_+}\frac{dz}{2\pi i}
\int_{C_-}\frac{dz^{\prime}}{2\pi i}f_1(z)f_2(z^{\prime})\delta(z,z^{\prime})=0,
\label{eq:C+C-=0}
\end{align}
which is obtained in the same way as that in \cite{Takahashi:2002ez} 
for the left/right half of the unit circle.

\section{Integration along $C_{\pm 1}^{s_{\rm max}}$
\label{sec:intCpm}}

We illustrate the evaluation of the integral along 
$C_{\pm 1}^{s_{\rm max}}$:
\begin{eqnarray}
 \int_{C_{\pm 1}^{s_{\rm max}}}\frac{dz}{2\pi i} g(z)f_\kappa(z)T(z).
 \label{eq:C+1smax}
\end{eqnarray}
Following the definition of $C_{\pm 1}^{s_{\rm max}}$ 
in \S  \ref{sec:virasoro}, 
we change the integration variable to the time $t$:
\begin{eqnarray}
\mbox{eq.~(\ref{eq:C+1smax})}
 &=&\int_{t\mp \varDelta t}^{t\pm \varDelta t}\frac{dt}{2\pi i}
g(z)^2 e^{\kappa (t\mp i s_{\rm max})}T(z),
\end{eqnarray}
where we have used $dz/dt=g(z)$ and $f_\kappa(z)=\exp(\kappa(t+is))$.

Suppose that the operator $T(z)$ is bounded from above in
correlation functions, i.e.,
$\big|\,\big<T(z)\cdots \big>\,\big|\leq M \big|\,\big<\cdots \big>\,\big|$.
Substituting (\ref{eq:gsin2}) for the above, we find
\begin{eqnarray}
 \left|\,\mbox{eq.~(\ref{eq:C+1smax})}\,\right|
&\leq & \int_{t-\varDelta t}^{t+\varDelta}\frac{dt}{2\pi}
\frac{e^{\kappa t}M}{\sqrt{(t^2+s_{\rm max}^2)^3\{
(t+2)^2+s_{\rm max}^2\}}}.
\label{eq:ineq_int}
\end{eqnarray}
Here we have used
\begin{eqnarray}
 g(z)^2=\frac{1}{(t+is)^3 (t+2+is)},
\end{eqnarray}
which is derived from $t+is =2/(z^2-1)$ and (\ref{eq:gsin2}).
From the mean value theorem, there exists $\xi$ in $(t-\varDelta t,\,
t+\varDelta t)$ such that
\begin{eqnarray}
\mbox{RHS of (\ref{eq:ineq_int})}
 &=&2\varDelta t\, \frac{1}{2\pi}
\frac{M e^{\kappa\xi}}{\sqrt{(\xi^2+s_{\rm max}^2)^3\{
(\xi+2)^2+s_{\rm max}^2\}}}.
\end{eqnarray}
By taking the limit $s_{\rm max}\rightarrow \infty$, 
the above approaches zero for a fixed value of $\varDelta t$.
Therefore, we obtain
\begin{eqnarray}
 \lim_{s_{\rm max}\rightarrow \infty}
  \int_{C_{\pm 1}^{s_{\rm max}}}\frac{dz}{2\pi i} g(z)f_\kappa(z)T(z)=0.
\end{eqnarray}
As a result, we can add the integral along ${C_{\pm 1}^{s_{\rm
max}}}$ to (\ref{eq:intTT}).

Here, we show that the essential singularity of $f_\kappa(z)$ has no
effect on the integration of the commutator of the Virasoro algebra.
The important point is that we quantize the system by the Hamiltonian
${\cal L}_0+\tilde{\cal L}_0$. So, for any weighting function $g(z)$,
$f_{\kappa}(z)$ is given by the differential equation (\ref{eq:deffk})
and it is written as
\begin{eqnarray}
 f_{\kappa}(z)=e^{\kappa(t+is)}.
\end{eqnarray}
On the $z$ plane, as a result of the quantization, 
each contour of string propagation
starts from and ends at the fixed points along the same direction. 
In the above example, it approaches $z=\pm 1$ parallel to
the imaginary axis. 
In general, a function takes the value of any complex
number in the neighborhood of its essential singularity. 
However, due to the direction of the contours, $f_\kappa(z)$
does not 
diverge if $z$ approaches the fixed point as $s_{\rm max}$ goes to
infinity  in the
integration. Hence, in general, the essential singularity in
$f_\kappa(z)$ becomes manageable by the quantization procedure of SSLD.

\section{Anti-commutation relations for the modified BRST operator
\label{sec:cr_modBRST}}

Expanding the BRST current $j_{\rm B}(z)=cT_{\rm mat}(z)+bc\partial c(z)+\frac{3}{2}\partial^2c(z)$
and 
the ghost $c(z)$ as
\begin{align}
 j_{{\rm B}}(z)=\sum_{n=-\infty}^{\infty}Q_{n}z^{-n-1},\qquad c(z)=\sum_{n=-\infty}^{\infty}c_{n}z^{-n+1},
 \label{eq:app_jB_c_mode}
 \end{align}
we have relations among the modes:
\begin{align}
\{Q_m,Q_n\}&=2mn\{Q_{\rm B},c_{m+n}\},
&\{Q_m,c_n\}&=\{Q_{\rm B},c_{m+n}\},
\end{align}
($Q_{\rm B}=Q_0$) and then the relations 
\begin{align}
\{j_{\rm B}(z),j_{\rm B}(z^{\prime})\}
&=2\partial\partial^{\prime}\left(\delta(z,z^{\prime})\{Q_{\rm B},c(z^{\prime})\}\right),
&\{j_{\rm B}(z),c(z^{\prime})\}&=\delta(z,z^{\prime})\{Q_{\rm B},c(z^{\prime})\}
\end{align}
are obtained in a similar way to the derivation in (\ref{eq:TT_proof}).
From these,  (\ref{eq:QpQp}), (\ref{eq:QpCp}) and (\ref{eq:QpQm})  are demonstrated
using (\ref{eq:C+C-=0}) and
\begin{align}
\int_{C_{\pm}}\frac{dz}{2\pi i}
\int_{C_{\pm}}\frac{dz^{\prime}}{2\pi i}f_1(z)f_2(z^{\prime})\delta(z,z^{\prime})=
\int_{C_{\pm}}\frac{dz}{2\pi i}f_1(z)f_2(z),
\label{eq:CpmCpm=Cpm}
\end{align}
which is derived similarly to (\ref{eq:C+C-=0}).
In particular, the conditions $f(\pm 1)=g(\pm 1)=0$ are used for 
$\{Q_{\pm}(f),Q_{\pm }(g)\}$ and $\{Q_{\pm}(f),Q_{\mp}(g)\}$.
For the operators $Q^{\prime}_{\pm}$ (\ref{eq:Q'pm_def}), 
we can compute them as
\begin{align}
\{Q_{\pm}^{\prime},Q_{\pm}^{\prime}\}&
=\{Q_{\pm}(e^h),Q_{\pm}(e^h)\}-2\{Q_{\pm}(e^h),C_{\pm}((\partial h)^2e^h)\}
\nonumber\\
&=2\{Q_{\rm B},C_{\pm}(\partial e^h)^2\}-2\{Q_{\rm B},C_{\pm}((\partial h)^2e^{2h}\}=0
\end{align}
with (\ref{eq:QpQp}) and (\ref{eq:QpCp})
because $e^{h(z)}$ vanishes at $z=\pm 1$ by assumption. Similarly, from (\ref{eq:QpQm}),
we have
\begin{align}
\{Q_{\pm}^{\prime},Q_{\mp}^{\prime}\}&
=\{Q_{\pm}(e^h),Q_{\mp}(e^h)\}-\{Q_{\pm}(e^h),C_{\mp}((\partial h)^2e^h)\}
-\{Q_{\mp}(e^h),C_{\pm}((\partial h)^2e^h)\}
=0.
\end{align}
Therefore, the nilpotency of $Q_{\pm}^{\prime}$ and anti-commutativity  (\ref{eq:nilpotent}) are proved.

\section{$Q^{\prime}_{\pm}$ invariance of $|\mathcal{V}\rangle_{\rm OSFT}$ 
\label{sec:QprimeInv} 
}

Here, we show  (\ref{eq:QpmV}), namely,
$Q^{\prime}_{\pm}$ invariance of $|\mathcal{V}\rangle_{\rm OSFT}$ (\ref{eq:calV}).
$Q_{\pm}^{\prime}$ (\ref{eq:Q'pm_def}) can be expressed as
\begin{align}
Q_{\pm}^{\prime}&=\int_{C_{\pm}}\frac{dz}{2\pi i}F(z)j_{{\rm B}}(z)+\int_{C_{\pm}}\frac{dz}{2\pi i}G(z)c(z),
\label{eq:Q'pm=FG}
\\
&F(z)=e^{h(z)},\qquad G(z)=-(\partial h(z))^{2}e^{h(z)}.
\end{align}
From the relation $h(-1/z)=h(z)$, we have
\begin{align}
&F(-1/z)=F(z),\qquad G(-1/z)=z^{4}G(z)
\end{align}
and therefore we can expand the weighting functions $F(z)$ and $G(z)$ as
\begin{align}
&F(z)=\sum_{n=0}^{\infty}a_{n}(z^{n}+(-1)^{n}z^{-n}),\qquad
G(z)=z^{-2}\sum_{n=0}^{\infty}a_{n}^{\prime}(z^{n}+(-1)^{n}z^{-n})
\end{align}
on the unit circle $|z|=1$, where $a_n$ and $a_n^{\prime}$ are constants.
Using the above and the mode expansions of the BRST current
and ghost (\ref{eq:app_jB_c_mode}),
(\ref{eq:Q'pm=FG}) can be rewritten as
\begin{align}
Q_{\pm}^{\prime} & =a_{0}Q_{0}+\sum_{n=1}^{\infty}\frac{a_{n}}{2}(Q_{n}+(-1)^{n}Q_{-n})+a_{0}^{\prime}c_{0}+\sum_{n=1}^{\infty}\frac{a_{n}^{\prime}}{2}(c_{n}+(-1)^{n}c_{-n})
\nonumber\\
 & \quad\mp\frac{i}{\pi}\Biggl[\sum_{k=0}^{\infty}\frac{2a_{0}}{2k+1}(Q_{2k+1}-Q_{-2k-1})+\sum_{k=0}^{\infty}\frac{2a_{0}^{\prime}}{2k+1}(c_{2k+1}-c_{-2k-1})
 \nonumber\\
 & \qquad\qquad+\sum_{n=1}^{\infty}\sum_{k=0}^{\infty}\frac{a_{n}}{2k+1}(Q_{n+2k+1}-(-1)^{n}Q_{-n-2k-1}-(Q_{n-2k-1}-(-1)^{n}Q_{-n+2k+1}))
 \nonumber\\
 & \qquad\qquad+\sum_{n=1}^{\infty}\sum_{k=0}^{\infty}\frac{a_{n}^{\prime}}{2k+1}(c_{n+2k+1}-(-1)^{n}c_{-n-2k-1}-(c_{n-2k-1}-(-1)^{n}c_{-n+2k+1}))\Biggr],
\end{align}
where use has been made of
\begin{align}
\int_{C_{\pm}}\frac{dz}{2\pi i}z^{-1-n} &  =\begin{cases}
\frac{1}{2} & (n=0)\\
0 & (n=2k\,,\ k\in\mathbb{Z}\setminus\!\{0\})\\
\frac{\mp i}{n\pi} & (n=2k+1\,,\ k\in\mathbb{Z}).
\end{cases}
\end{align}
In particular, we find that $Q^{\prime}_{\pm}$ is given by a linear combination of 
\begin{align}
Q_{n}+(-1)^{n}Q_{-n}\,,\qquad c_{n}+(-1)^{n}c_{-n}\,,\qquad(n\in\mathbb{Z}),
\end{align}
and hence we have (\ref{eq:QpmV}),
thanks to
\begin{align}
 &\left(Q_{n}+(-1)^{n}Q_{-n}\right) |\mathcal{V}\rangle_{\rm OSFT}=0,\qquad\left(c_{n}+(-1)^{n}c_{-n}\right) |\mathcal{V}\rangle_{\rm OSFT}=0,
\end{align}
which can be shown as in \cite{Kawano:2008ry} using a method in \cite{Rastelli:2000iu}.

\section{Commutation relations for the deformed Hamiltonian at the
  tachyon vacuum
\label{sec:cr_tv}}
For the holomorphic and antiholomorphic parts of the Hamiltonian, $H_{\pm}$ (\ref{eq:Htachyon}),
commutation relations involving the ghost number current $j_{\rm gh}(z)=cb(z)=\sum_{n\in\mathbb{Z}}q_nz^{-n-1}$ are necessary. 
From the OPEs, we have
\begin{align}
 [L_m,q_n]&=-nq_{m+n}-\frac{3}{2}m(m+1)\delta_{m+n,0},
&[q_m,q_n]&=m\delta_{m+n,0},
\end{align}
and then
\begin{align}
  [T(z),j_{\rm gh}(z^{\prime})]&=\partial^{\prime}\delta(z,z^{\prime})j_{\rm gh}(z)
 -\frac{3}{2}\partial^2\delta(z,z^{\prime}),
 &[j_{\rm gh}(z),j_{\rm gh}(z^{\prime})]&=-\partial\delta(z,z^{\prime}),
 \label{eq:Tjgh_jghjgh}
\end{align}
are derived in a similar way to (\ref{eq:TT_proof}).
Using (\ref{eq:TT}), (\ref{eq:Tjgh_jghjgh}) and (\ref{eq:C+C-=0}), we have
\begin{align}
 &[H_+,H_-]\nonumber\\
 &=\int_{C_+}\frac{dz}{2\pi i}\int_{C_-}\frac {dz^{\prime}}{2\pi i}zz^{\prime}
  [e^hT(z)+\partial e^h j_{\rm gh}(z),e^hT(z^{\prime})+\partial e^h j_{\rm gh}(z^{\prime})]
  \nonumber\\
  &=\int_{C_+}\frac{dz}{2\pi i}\int_{C_-}\frac {dz^{\prime}}{2\pi i}
 zz^{\prime}\biggl( 
 -e^{h(z)}e^{h(z^{\prime})}
 \left(
 (T(z)+T(z^{\prime}))\partial\delta(z,z^{\prime})
+\frac{c}{12}\partial^3\delta(z,z^{\prime})
 \right)
 -\partial e^{h(z)}\partial e^{h(z^{\prime})}\partial\delta(z,z^{\prime})
 \nonumber\\
 &\qquad+e^{h(z)}\partial e^{h(z^{\prime})}\left(
 \partial^{\prime}\delta(z,z^{\prime})j_{\rm gh}(z)
 -\frac{3}{2}\partial^2\delta(z,z^{\prime})\right)
 -\partial e^{h(z)}e^{h(z^{\prime})}\left(
\partial\delta(z,z^{\prime})j_{\rm gh}(z^{\prime})
 -\frac{3}{2}\partial^{\prime 2}\delta(z,z^{\prime})\right)
 \biggr)
 \nonumber\\
  &=\int_{C_+}\frac{dz}{2\pi i}\int_{C_-}\frac {dz^{\prime}}{2\pi i}
 \biggl( 
 \partial(z e^{h})z^{\prime}e^{h(z^{\prime})}
 (T(z)+T(z^{\prime}))
-\frac{c}{12}
 \partial^2(z e^{h})\partial^{\prime}(z^{\prime}e^{h})
 +\partial(z\partial e^{h})z^{\prime}\partial^{\prime} e^{h}
 \nonumber\\
 &\qquad-ze^{h}j_{\rm gh}(z)\partial^{\prime}(z^{\prime}\partial^{\prime} e^{h})
 -\frac{3}{2}\partial^2(ze^{h})z^{\prime}\partial^{\prime} e^{h}
 +\partial(z\partial e^{h})z^{\prime}e^{h}j_{\rm gh}(z^{\prime})
 +\frac{3}{2}z\partial e^{h}\partial^{\prime 2}(z^{\prime} e^{h})
 \biggr)\delta(z,z^{\prime})
 \nonumber\\
 &=0,
\end{align}
because $e^{h(z)}$ has second-order zeros at $z=\pm 1$ by assumption.
Actually, the central charge for $T(z)$ is zero in the context of (\ref{eq:Htachyon}).

\section{BRST operators, Virasoro operators and ghost number at the
tachyon vacuum
\label{sec:BRST_tv}}

The BRST current $j_{\rm B}(z)$ 
and the ghost $c(z)$ are primary fields with conformal weights $1$ and $-1$, respectively, 
and hence we have
\begin{align}
 [Q_m,L_n]&=mQ_{m+n},
 &[c_m,L_n]&=(2n+m)c_{m+n}
\end{align}
which imply the commutation relations
\begin{align}
 [j_{\rm B}(z),T(z^{\prime})]&=-\partial\delta(z,z^{\prime})j_{\rm B}(z^{\prime}),
 &[c(z),T(z^{\prime})]&=\partial\delta(z,z^{\prime})c(z)-2\delta(z,z^{\prime})\partial c(z^{\prime}).
 \label{eq:jBT_cT}
\end{align}
For the ghost number current $j_{\rm gh}(z)$,
the modes satisfy the commutation relations
\begin{align}
 [Q_m,q_n]&=-Q_{m+n}+2mnc_{m+n},
&[c_m,q_n]&=-c_{m+n},
\end{align}
and they give
\begin{align}
 [j_{\rm B}(z),j_{\rm gh}(z^{\prime})]&=-\delta(z,z^{\prime})j_{\rm B}(z^{\prime})+2\partial\partial^{\prime}
 \left(\delta(z,z^{\prime})c(z^{\prime})\right),
&[c(z),j_{\rm gh}(z^{\prime})]&=-\delta(z,z^{\prime})c(z^{\prime}).
\label{eq:jBjgh_cjgh}
\end{align}
If we define $j^{\prime}_{\rm B}(z)$ 
such as $Q^{\prime}_{\pm }=\int_{C_{\pm}}\frac{dz}{2\pi i}j^{\prime}_{\rm B}(z)$ by
\begin{align}
j^{\prime}_{\rm B}(z)&=e^{h(z)}j_{\rm B}(z)-(\partial h(z))^2e^{h(z)}c(z),
\label{eq:j_Bprime_def}
\end{align}
we can compute a commutation relation with ${\mathcal T}(z)$ (\ref{eq:calT})
using (\ref{eq:jBT_cT}) and (\ref{eq:jBjgh_cjgh}) as
\begin{align}
 [j^{\prime}_{\rm B}(z),{\mathcal T}(z^{\prime})]&=-\partial
 \left(e^{h(z)}\delta(z,z^{\prime})j_{\rm B}(z^{\prime})\right)
 \nonumber\\
 &\quad +\partial\left(
 2e^{h(z)}\partial h(z^{\prime})\partial^{\prime}(\delta(z,z^{\prime})c(z^{\prime}))
 +e^{-h(z^{\prime})}(\partial e^{h(z^{\prime})})^2\delta(z,z^{\prime})c(z^{\prime})
 \right).
\end{align} 
Integrating the above, we have
\begin{align}
 [Q^{\prime}_{\pm},g(z^{\prime})f_{\kappa}(z^{\prime}){\mathcal T}(z^{\prime})]
 &=\mp \left[z^{\prime}e^{h(z)}e^{h(z^{\prime})}
 e^{\kappa w^{\prime}}\delta(z,z^{\prime})j_{\rm B}(z^{\prime})
 \right]^{z=-1}_{z=1}
 \nonumber\\
 &\quad\pm \left[
 2z^{\prime}e^{h(z)}\partial e^{h(z^{\prime})}
 e^{\kappa w^{\prime}}
 \partial^{\prime}(\delta(z,z^{\prime})c(z^{\prime}))
 +z^{\prime}(\partial e^{h(z^{\prime})})^2e^{\kappa w^{\prime}}\delta(z,z^{\prime})c(z^{\prime})
 \right]^{z=-1}_{z=1}
 \label{eq:QpgfkT}
\end{align}
for $g(z^{\prime})=z^{\prime}e^{h(z^{\prime})}$ and $f_{\kappa}(z^{\prime})=e^{\kappa w^{\prime}}$
with
$w^{\prime}=\int^{z^{\prime}}\frac{dz}{g(z)}$.
Noting that $e^{h(z)}$ has second-order zeros at $z=\pm 1$ by
assumption and 
assuming that $| e^{\kappa w^{\prime}}|$ is bounded from above 
near $z^{\prime}=\pm 1$ along the unit circle,\footnote{
In the case of (\ref{eq:gsin2}), we have $w^{\prime}=\frac{2}{z^{\prime 2}-1}=-1-i\cot\theta$
for $z^{\prime}=e^{i\theta}$ and hence $| e^{\kappa w^{\prime}}|=e^{-\kappa}$
for a real number $\kappa$.
} 
we obtain (\ref{eq:3.23}) by integrating (\ref{eq:QpgfkT}) along $C_{\pm}$.

From (\ref{eq:jBjgh_cjgh}), we have a commutation relation:
\begin{align}
 [j_{\rm B}^{\prime}(z), j_{\rm gh}(z^{\prime})]&=-\delta(z,z^{\prime})j^{\prime}_{\rm B}(z^{\prime})+\partial^{\prime}\left(2e^{h(z)}\partial\delta(z,z^{\prime})c(z^{\prime})
 \right).
\end{align}
Integrating the above and using (\ref{eq:CpmCpm=Cpm}),
we have
\begin{align}
 \int_{C_{\pm}}\frac{dz}{2\pi i}\int_{C_{\pm}}\frac{dz^{\prime}}{2\pi i}[j_{\rm B}^{\prime}(z), j_{\rm gh}(z^{\prime})]&=- \int_{C_{\pm}}\frac{dz}{2\pi i}j^{\prime}_{\rm B}(z)
 \mp\left[
 \int_{C_{\pm}}\frac{dz}{2\pi i}
 2(\partial e^{h(z)})\delta(z,z^{\prime})c(z^{\prime})
 \right]^{z^{\prime}=-1}_{z^{\prime}=1}.
 \end{align}
Because $\partial e^{h(z)}$ vanishes at $z=\pm 1$, 
the second term on the right-hand side is zero.
Similarly, from  (\ref{eq:C+C-=0}), we have
\begin{align}
  \int_{C_{\pm}}\frac{dz}{2\pi i}\int_{C_{\mp}}\frac{dz^{\prime}}{2\pi i}[j_{\rm B}^{\prime}(z), j_{\rm gh}(z^{\prime})]&=0.
 \end{align}
These relations imply (\ref{eq:3.32}).

On the modified BRST current, 
we note that the difference of  (\ref{eq:3.38}) and (\ref{eq:j_Bprime_def}) 
is computed as
\begin{align}
{\mathcal J}_{\rm B}(z)-j^{\prime}_{\rm B}(z)&=\partial\left(
2(\partial e^{h(z)}) c(z)\right)
\label{eq:app_calJB-jB'}
\end{align}
and therefore (\ref{eq:3.39}) holds.

It is possible to provide an alternative derivation of
(\ref{eq:3.23}). 
First, we introduce continuous modes for the anti-ghost
field at the tachyon vacuum:
\begin{eqnarray}
 b_\kappa^{\prime} =\int_{C_+}\frac{dz}{2\pi i}g(z)f_\kappa(z)b'(z),
~~~~
 \tilde{b}_\kappa^{\prime} =\int_{C_-}\frac{dz}{2\pi i}g(z)f_\kappa(z)b'(z).
 \label{eq:bkappa_def}
\end{eqnarray}
Since ${\cal J}_{\rm B}(z)$, $b'(z)$, ${\cal T}(z)$ and ${\mathcal J}_{\rm gh}(z)$ satisfy the same
OPE as the conventional ones, we find the anti-commutation relation\footnote{
We note that the conventional modes satisfy
$
\{Q_m,b_m\}=L_{m+n}+mq_{m+n}+\frac{3}{2}m(m-1)\delta_{m+n,0}
$,
which can be derived from the OPE for $j_{\rm B}(z)b(z^{\prime})$.
}
\begin{align}
 \{{\mathcal J}_{\rm B}(z),b^{\prime}(z^{\prime})\}
 &=\delta(z,z^{\prime}){\mathcal T}(z^{\prime})-\partial\delta(z,z^{\prime}){\mathcal J}_{\rm gh}(z^{\prime})
 +\frac{3}{2}\partial^2\delta(z,z^{\prime}).
\end{align}
Integrating the above, we have
\begin{align}
\{Q^{\prime}_{\pm},b_{\kappa}^{\prime}\}&=
\int_{C_{\pm}}\frac{dz}{2\pi i}\int_{C_{+}}\frac{dz^{\prime}}{2\pi i}
g(z^{\prime})f_{\kappa}(z^{\prime}){\mathcal T}(z^{\prime})\delta(z,z^{\prime})
\nonumber\\
&\qquad \mp \left[
\int_{C_{+}}\frac{dz^{\prime}}{2\pi i}\left(
g(z^{\prime})f_{\kappa}(z^{\prime}){\mathcal J}_{\rm gh}(z^{\prime})
+\frac{3}{2}(\partial g(z^{\prime})+\kappa)
f_{\kappa}(z^{\prime})
\right)\delta(z,z^{\prime})
\right]^{z=-1}_{z=1},
\label{eq:Q'pmb'kappa}
\end{align}
where we have assumed that $|f_{\kappa}(z^{\prime})|$ 
is bounded from above near $z^{\prime}=\pm 1$ along the unit circle.
The second line on the right-hand side of (\ref{eq:Q'pmb'kappa}) 
does not contribute by taking the same regularization as (\ref{eq:Ldef}) 
in (\ref{eq:bkappa_def}) 
and then we obtain
\begin{align}
\{Q'_+,\,b_\kappa^{\prime}\}&={\mathcal L}_{\kappa},
&\{Q'_-,\,b_\kappa^{\prime}\}&=0
\end{align}
from (\ref{eq:CpmCpm=Cpm}), (\ref{eq:C+C-=0}) and (\ref{eq:calLkappa_def}).
Similarly, 
\begin{align}
\{Q'_-,\,{\tilde b}_\kappa^{\prime}\}&={\tilde {\mathcal L}}_{\kappa},
&\{Q'_+,\,{\tilde b}_\kappa^{\prime}\}&=0
\end{align}
are obtained. With the above anti-commutation relations,  (\ref{eq:nilpotent}) and the Jacobi identity,
eq.~(\ref{eq:3.23}) is derived.


\end{document}